\documentclass[%
 reprint,
superscriptaddress,
 amsmath,amssymb,
 aps,
]{revtex4-2}

\usepackage{graphicx}
\usepackage{dcolumn}
\usepackage{bm}
\usepackage{braket}
\usepackage{color}
\usepackage{textcomp}

\usepackage{amsmath,amsfonts}
\usepackage{mathrsfs}

\begin{document}

\preprint{APS/123-QED}

\title{Super spontaneous four-wave mixing in an array of silicon microresonators}

\author{Massimo Borghi}
\email{Corresponding authors: massimo.borghi@unipv.it -  matteo.galli@unipv.it}
\affiliation{Dipartimento di Fisica, Universit\`a di Pavia, Via Agostino Bassi 6, 27100 Pavia, Italy}
\author{Federico Andrea Sabattoli}\email{Current address: Advanced Fiber Resources Milan S.r.L, via Fellini 4, 20097 San Donato Milanese (MI), Italy}
\affiliation{Dipartimento di Fisica, Universit\`a di Pavia, Via Agostino Bassi 6, 27100 Pavia, Italy}

\author{Houssein El Dirani}
\email{Current address: LIGENTEC SA, 224 Bd John Kennedy, 91100 Corbeil-Essonnes, France}
\affiliation{Univ. Grenoble Alpes, CEA-Leti, 38054 Grenoble cedex, France}

\author{Laurene Youssef} \email{Current address: IRCER,  Centre Européen de la Céramique, 12 rue Atlantis, 87068 Limoges, France}
\affiliation{Univ. Grenoble Alpes, CNRS, LTM, 38000 Grenoble, France}



\author{Camille Petit-Etienne}
\affiliation{Univ. Grenoble Alpes, CNRS, LTM, 38000 Grenoble, France}

\author{Erwine Pargon}
\affiliation{Univ. Grenoble Alpes, CNRS, LTM, 38000 Grenoble, France}

\author{
\textcolor{black}{J.}E. Sipe}
\affiliation{Department of Physics, University of Toronto, 60 St. George Street, Toronto, ON, M5S 1A7, Canada}
\author{Amideddin Mataji-Kojouri}
\affiliation{Dipartimento di Fisica, Universit\`a di Pavia, Via Agostino Bassi 6, 27100 Pavia, Italy}

\author{Marco Liscidini}
\affiliation{Dipartimento di Fisica, Universit\`a di Pavia, Via Agostino Bassi 6, 27100 Pavia, Italy}

\author{Corrado Sciancalepore}\email{Current address: SOITEC SA, Parc technologique des Fontaines, Chemin des Franques, 38190 Bernin, France}
\affiliation{Univ. Grenoble Alpes, CEA-Leti, 38054 Grenoble cedex, France}

\author{Matteo Galli} 
\affiliation{Dipartimento di Fisica, Universit\`a di Pavia, Via Agostino Bassi 6, 27100 Pavia, Italy}

\author{Daniele Bajoni}
\affiliation{Dipartimento di Ingegneria Industriale e dell'Informazione, Universit\`a di Pavia, Via Adolfo Ferrata 5, 27100 Pavia, Italy}

%




\date{\today}

\begin{abstract}
\noindent  Composite \textcolor{black}{optical} 
systems can show compelling collective dynamics. 
For instance, the cooperative decay of quantum emitters into a common radiation mode can lead to superradiance, where the emission rate of the ensemble is larger than the sum of the rates of the individual emitters. Here, we report experimental evidence of super spontaneous four-wave mixing (super SFWM),  an analogous effect for the generation of photon pairs in a parametric nonlinear process on an integrated photonic device. We study this phenomenon in an array of 
\textcolor{black}{microring resonators} 
on a silicon photonic chip \textcolor{black}{coupled to bus waveguides}. We measured a cooperative pair generation rate that always exceeds the incoherent sum of the rates of the individual resonators. We investigate the physical mechanisms underlying this collective behaviour, 
clarify the impact of loss, and address the aspects of fundamental and technological relevance of our results.

\end{abstract}

\maketitle


\section{Introduction}
\noindent When multiple 
quantum emitters are coupled to 
the electromagnetic field, their collective 
\textcolor{black}{dynamics}  can fundamentally differ from 
\textcolor{black}{that of isolated emitters.} \textcolor{black}{``}Superradiance\textcolor{black}{,"} which is \textcolor{black}{the enhancement of the} 
spontaneous emission \textcolor{black}{rate of} 
a collection of 
\textcolor{black}{such emitters,} is 
\textcolor{black}{a quintessential example of this difference.} \textcolor{black}{The pioneering work on superradiance was due to }
Dicke \cite{dicke1954coherence}, \textcolor{black}{and since his paper in 1954} the effect has been \textcolor{black}{the} subject of a large number of theoretical investigations and experimental demonstrations (see, e.g., \cite{gross1982superradiance,scully2009super,benedict2018super} and references therein),  
 acquiring technological relevance in the fields of quantum information science \cite{gonzalez2015deterministic,asenjo2017exponential,facchinetti2016storing}, laser physics \cite{bohnet2012steady} and metrology \cite{paulisch2019quantum}. Although conceptually simple,  its experimental validation is challenged by \textcolor{black}{difficulties in precisely preparing} 
 the initial quantum state, and \textcolor{black}{in controlling detrimental interactions with the environment} 
 \cite{gross1982superradiance,scully2009super}. In experiments involving dilute atomic clouds, this is often achieved by laser cooling in magneto-optical traps \cite{araujo2016superradiance,skribanowitz1973observation,srivathsan2013narrow}. The presence of a large number of atoms only provides a macroscopic description of the behaviour of the system. 
Yet, \textcolor{black}{with progress in materials science and nanotechnology, it is now possible to} 
initialize a collection of single emitters 
\textcolor{black}{in} the so-called Dicke bright and dark states \textcolor{black}{ -- } i.e., the maximally super- and subradiant states \textcolor{black}{ -- with high fidelity} \cite{gonzalez2015deterministic}. 
Moreover, selective coupling to  nanophotonic waveguides or cavities can be engineered to tailor the spontaneous emission into a restricted set of radiation modes \cite{goban2015superradiance,lukin2022optical}. 
Recent studies have 
demonstrated cooperative effects between quantum emitters 
\textcolor{black}{involving} a wide variety of platforms, including superconducting qubits \cite{wang2020controllable,mlynek2014observation}, quantum dots \cite{kim2018super,koong2022coherence,grim2019scalable}, trapped ions \cite{goban2015superradiance}, cold atoms \cite{guerin2016subradiance}, and colour centers in diamond \cite{sipahigil2016integrated,angerer2018superradiant}. \textcolor{black}{These platforms need to operate in cryogenic environments, for the systems under study 
are notoriously prone to decoherence.} 
\begin{figure*}[t!]
\centering\includegraphics[scale = 0.74]{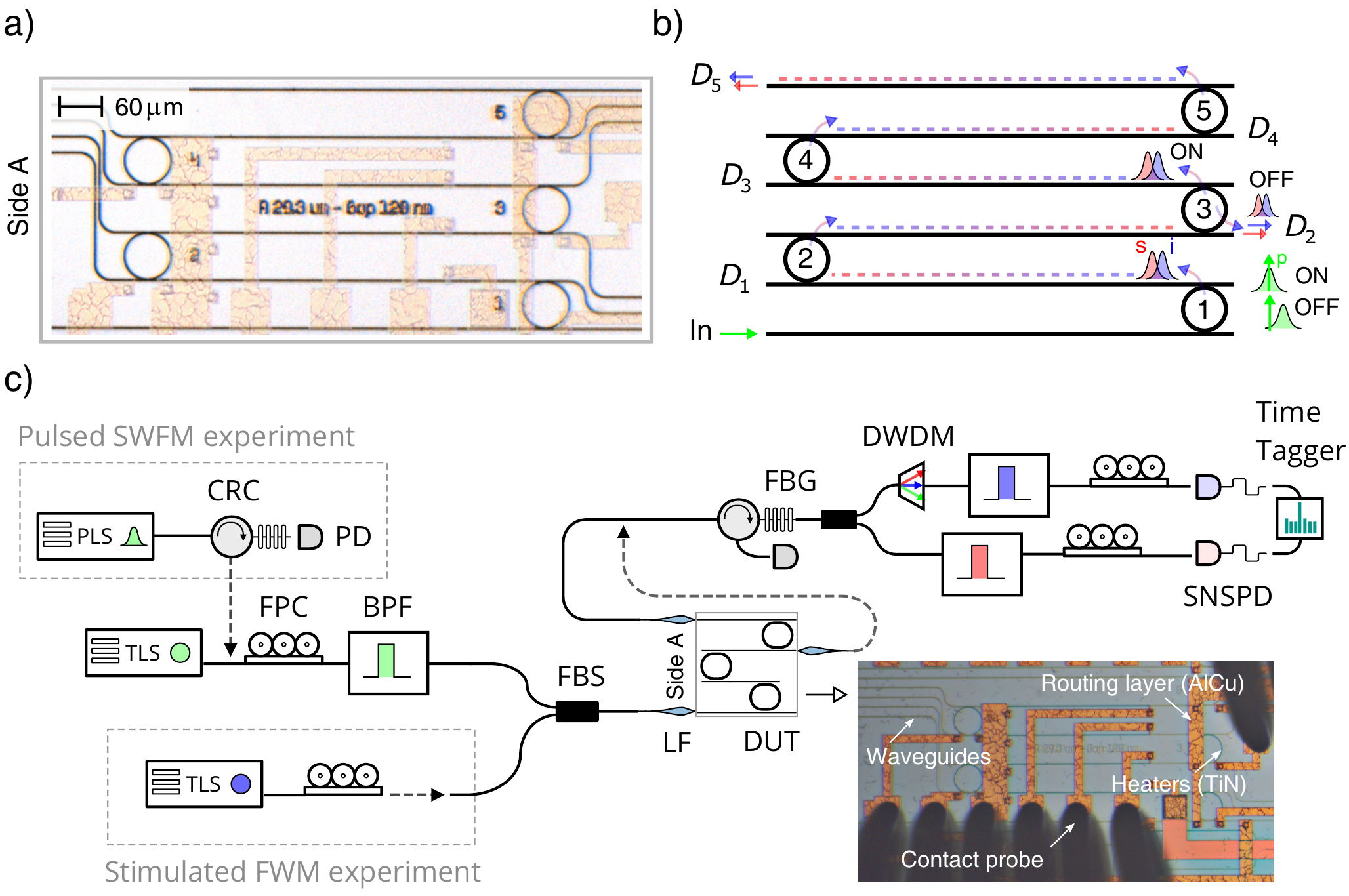}
\caption{(a) Optical microscope image of the device. The metallic routing layer realized in AlCu (shown in gold) is overlaid to that of the waveguides (black) for better clarity. (b) Principle of operation. The pump light (green) is injected at the input (In), and can be routed from the first $(1)$ to the last $(5)$ resonator. A ring in a ON state resonantly drops the pump and the photon pairs (s=signal, i=idler, shown respectively in red and blue) towards the next resonator in the sequence. In the OFF state, the ring is tuned out of resonance, and both the pump and the signal/idler pairs exit from the bus waveguide. (c) Sketch of the experimental setup used to perform both spontaneous and stimulated FWM on the Device Under Test (DUT). PLS = Pulsed Laser Source, TLS = Tunable Laser Source, BPF = Band Pass Filter, FPC = Fiber Polarization Controller, LF = Lensed Fiber, FBG = Fiber Bragg Grating, DWDM = Dense Wavelength Demultiplexing Module, SNSPD = Superconducting Nanowire Single Photon Detector.
\label{Fig_1}}
\end {figure*}
Superradiance in parametric processes such as four-wave mixing \cite{srivathsan2013narrow} has been subject of several works, mostly focusing on coherent matter-wave amplification \cite{scully2006directed,schneble2003onset}.

\noindent Recently, Onodera et al. \cite{onodera2016parametric} have \textcolor{black}{identified} 
\textcolor{black}{a} 
striking resemblance between Dicke superradiance and the enhancement of parametric fluorescence from an ensemble of identical resonators. 
In their theoretical work, the authors considered 
spontaneous four-wave mixing (SFWM) in a sequence of lossless microring resonators \textcolor{black}{arranged} in an all-pass configuration, and named the process super SFWM. 
\textcolor{black}{For} a quasi continuous-wave (CW) pump, and in the limit of a small probability of generating a photon pair within the pump coherence time, \textcolor{black}{they predicted that} the pair generation rate 
\textcolor{black}{would be} characterized by a super-linear scaling with the number $N$ of resonators. 
\textcolor{black} {Superradiance here is associated with} 
the impossibility of knowing in which resonator the photon pair is generated, with 
\textcolor{black}{the quantum state} given by a coherent superposition of the states associated with a pair being generated in \textcolor{black}{each} 
of the resonators. The analogy was drawn completely in the optical domain and on an integrated photonic device, considerably simplifying its experimental realization. Beside the fundamental aspect, the demonstration of super SFWM would acquire technological relevance for the realization of bright sources based on cooperative photon pair emission between resonators.
\textcolor{black}{Yet,} the theoretical work of Onodera et. al \cite{onodera2016parametric} did not consider the effects of propagation losses, unavoidable in any actual implementation. Indeed, in the all-pass geometry considered by the authors, such effect would have progressively decreased the transmittance along the sequence, spoiling very soon the cooperative character of spontaneous emission. 
Would it be possible to observe super SFWM in the presence of losses? What would it be their impact? 

\noindent We addressed those questions by experimentally investigating super SFWM in an array of ring resonator photon pair sources \cite{llewellyn2020chip} integrated on a silicon chip. We exploit the integrated photonic platform to 
control \textcolor{black}{precisely} the state of each emitter, harnessing the reconfigurability of the device to vary the number of collectively excited emitters. 
We study the scaling of the efficiencies of spontaneous and stimulated FWM \cite{azzini2012classical} for an increasing \textcolor{black}{array} size, 
\textcolor{black}{demonstrating} 
profound differences with respect to the case of an ensemble of independent resonators. 
This \textcolor{black}{allows us to prove that the array exhibits cooperative emission.} 
The impact of loss and spectral filtering on the pump and on the photon pairs \textcolor{black}{generated} are investigated 
\textcolor{black}{both experimentally and by numerical simulation.} 
\begin{figure}[t!]
\centering\includegraphics[scale = 0.465]{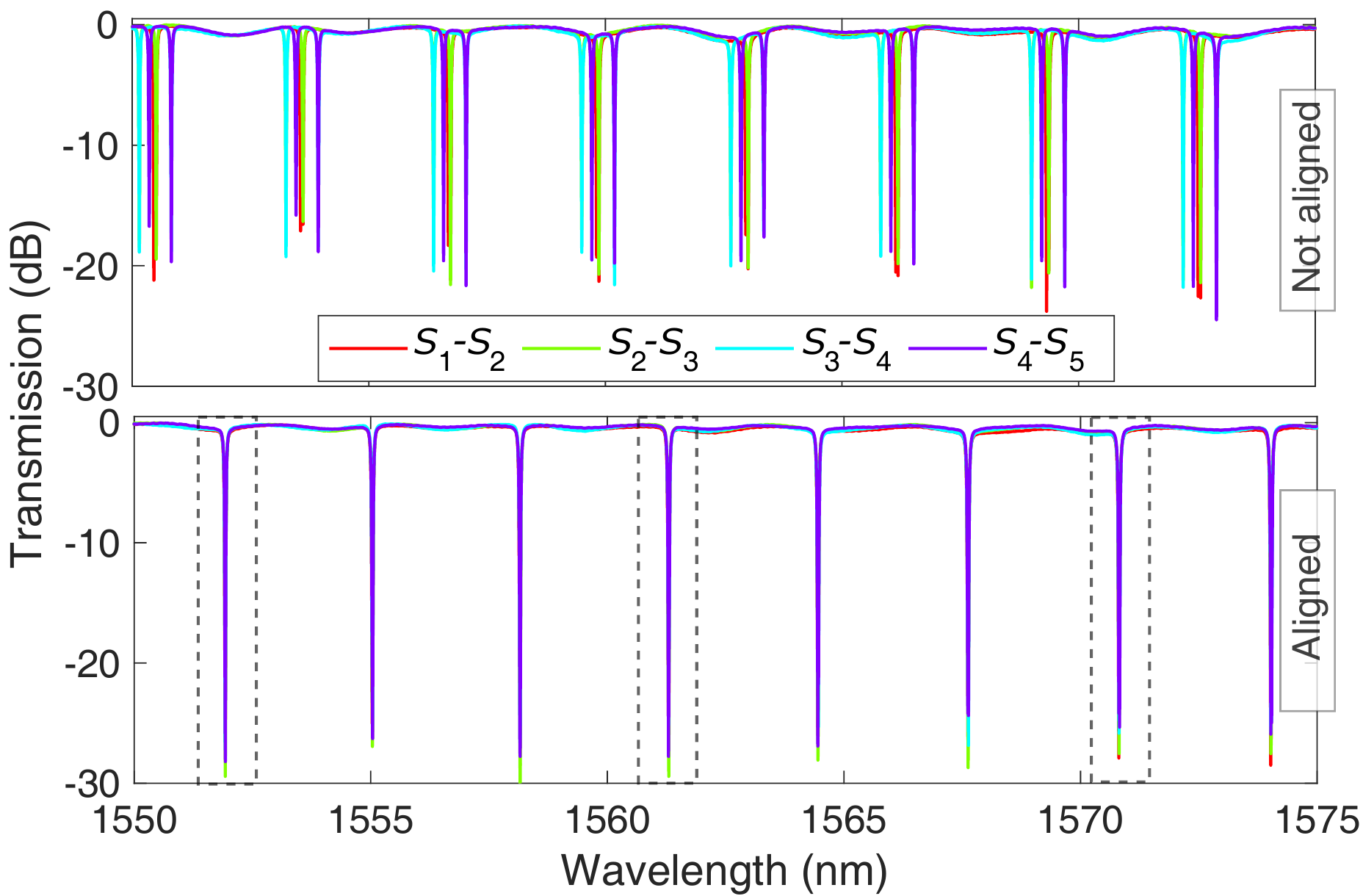}
\caption{Transmission spectra of the resonator sources ($S_1-S_5$) when no thermal tuning is applied (upper panel), and when the spectra of all the rings are overlapped (lower panel). From left to right, the idler, the pump and signal resonance order are marked in a dashed box.
\label{Fig_2}}
\end {figure}
\section{Device and experimental setup}
An optical microscope image and a functional scheme of the device are shown in Fig.\ref{Fig_1}(a) and Fig.\ref{Fig_1}(b), respectively. The photonic circuit consists of an array of five ring resonators (emitters) in the add-drop filter configuration \cite{carolan2019scalable}, where each drop port is used to excite the next emitter in the sequence. We label the drop port of the $j_{\text{th}}$ ring as $D_j$, as indicated in  Fig.\ref{Fig_1}(b). Contrarily to the all-pass scheme proposed in \cite{onodera2016parametric}, this configuration avoids the drop of the pump power, as well of photon pairs, along the sequence. At the same time, it does not compromise the possibility of coherently exciting all the rings simultaneously. The device is patterned on a 
\textcolor{black}{Silicon-on-Insulator} wafer using standard photolithography (details in the Supplemental Material \cite{suppl}).
The experimental setup is shown in Fig.\ref{Fig_1}(c). The chip is mounted on a sample holder that is temperature stabilized. Metallic micro-heaters allow \textcolor{black}{for the independent tuning of the resonance wavelength of each} 
ring through the thermo-optic effect. The applied electrical current is controlled by two separate multi-contact probes (bottom and top of the inset in Fig.\ref{Fig_1}(c)) which are placed in contact to the chip metallic pads. The probes are connected to five independent current/voltage output channels. The chip is designed to have the drop ports $D_j$, with $j$ odd, on the left-hand side (side A in the inset of Fig.\ref{Fig_1}), while the even ports lie on the right-hand side.
Micrometric positioning stages are used to align lensed fibers to the input/output chip waveguides.
 Two fibers are placed on side A, and are respectively used to inject the pump light and to collect photon pairs from the ports $D_1,D_3$ and $D_5$. The difference of the transmission losses between the fibers that are used to collect photon pairs is about $\sim 0.3$ dB. This offset has been applied to calibrate the coincidence rates collected from the even and the odd drop ports. A coupling loss of about $\sim 2.4$ dB/facet is achieved using waveguide terminated inverse tapers of width around $\sim130$ nm. 
With the heaters off, we injected light into ports $D_1$ through $D_4$ and probed the transmission spectra of the different rings with a CW tunable laser.  These   are shown superimposed in the upper panel of Fig.\ref{Fig_2}. The rings \textcolor{black}{possess} slightly different resonance wavelengths, showing in the spectra as clustered dips which are separated by multiples of the resonator free spectral range (FSR). 
By regulating the current in the heaters, the spectra of the rings can be overlapped, which is the condition shown in the lower panel of Fig.\ref{Fig_2}. 
In this configuration, the pump light, \textcolor{black}{injected at the input and on resonance with all the rings,} 
is dropped multiple times (average loss of $0.88\pm0.07$ dB for each drop event) and coherently excites all the emitters in the array. The path from the input to $D_N$ (dashed line in Fig.\ref{Fig_1}(b)) defines a common channel, shared by all the resonators, 
\textcolor{black}{and in which} 
it is not possible to discern which emitter fired. We can reconfigure the length of the sequence to $K<N$ by setting the  ring $K+1$ out of resonance (OFF state in Fig.\ref{Fig_1}(b)).  
Signal and idler photons generated by SFWM are collected from two resonances at $\lambda_s= 1571.2\,\textrm{nm}$ and $\lambda_i = 1551.4\,\textrm{nm}$, which  are separated by two FSR from the pump wavelength $\lambda_p = 1561.25\,\textrm{nm}$. The average quality factors are $Q_p = (3.9\pm0.2)\times 10^4$, $Q_s = (3.7\pm0.1)\times 10^4$ and $Q_i = (4.2\pm0.3)\times 10^4$, showing good uniformity among the rings (see Supplemental Material \cite{suppl} for a comprehensive characterization).
The collected pairs undergo a preliminar filtering stage which attenuates ($\sim 30$ dB) the pump beam through a Bragg grating and a circulator. Then, they are probabilistically separated using a $50/50$ fiber beam splitter, and the signal/idler paths are defined by the different center wavelengths of two sets of bandpass filters, achieving nearly $80$ dB of pump suppression. Their bandwidth is sufficiently large to accommodate two resonance orders within their passband. In order to select the coincidence events originating from only one set of resonances, we placed an additional $100$ GHz Dense Wavelength Demultiplexing Module (DWDM) on the idler path. Coincidence measurements are performed using two superconducting detectors and 
time tagging electronics. 
\begin{figure}[h!]
\centering\includegraphics[scale = 0.61]{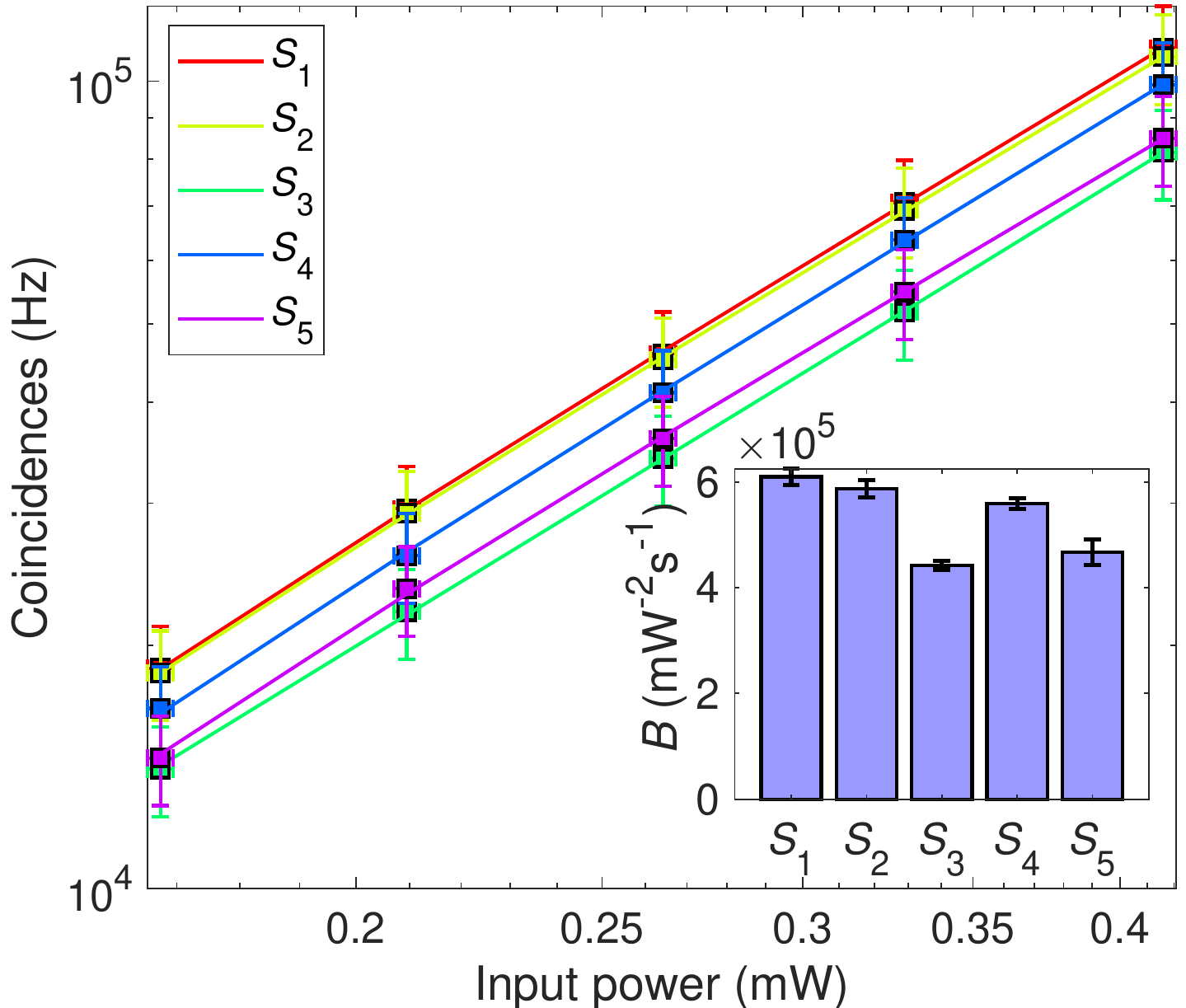}
\caption{Coincidences between the signal and idler photons as a function of the pump power coupled to the input waveguide. Scatters are experimental data, while solid lines are a quadratic fit. 
The average slope of the curves, in a double logarithmic scale, is $1.94\pm0.02$. The inset shows the brightness B ($\textrm{mW}^{-2}s^{-1}$) of each source. 
\label{Fig_3}}
\end{figure}
\begin{figure}[t!]
\centering\includegraphics[scale = 0.44]{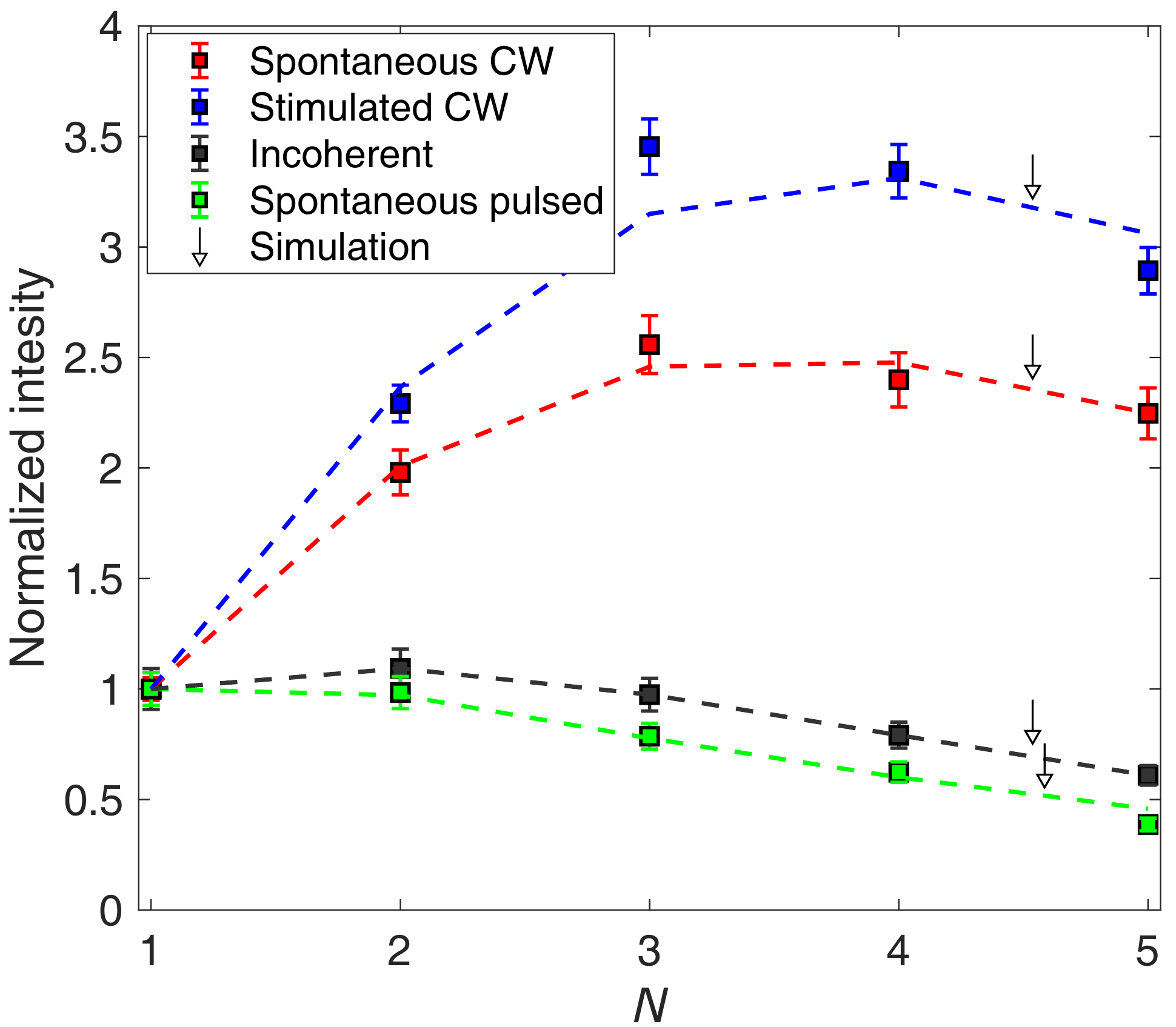}
\caption{Scaling of the intensity of different processes as a function of the size of the array of resonators $N$. In case of SFWM (red for CW excitation, green for pulsed pump), the intensity corresponds to the coincidence rate $R(N)$ measured at the output port $D_N$, normalized by the rate of the first ring $R(1)$. In stimulated FWM (blue), the intensity is relative to the signal beam at $1571.2$ nm. The incoherent scaling (black) is obtained by pumping each ring individually, 
as described in the main text. The dashed lines indicated by arrows represent numeric simulations. \label{Fig_4}}
\end{figure}
\section{Results}
As a first step, we characterized the brightness $B_j$ ($j=1,..,5$) of each source $S_j$ to evaluate the uniformity of the array. To test $S_j$, we coupled the pump laser to port $D_j$, and tuned the sources $S_j\neq S_k$ into an OFF state. Photon pairs are collected in transmission as a function of the input power. The coincidence rate at the output of the resonator are reported in Fig.\ref{Fig_3}. All the curves show the characteristic quadratic scaling of SFWM with the pump power. The brightness of the emitters is quite uniform, with an average value of $(5.3\pm0.7)\times 10^{5}\,\textrm{pairs}/(\textrm{s}\cdot\textrm{mW}^2)$. 

We then investigated the cooperative emission of the ring array by measuring the coincidence rate as a function of the  number $N$ of coherently pumped resonators. 
\textcolor{black}{To do this,} we excited the input of the sequence and overlapped the spectra of the emitters from $S_1$ to $S_{N}$, collecting pairs from port $D_N$. The coincidence rates versus $N$, normalized with respect to that of source $S_1$, are reported in Fig.\ref{Fig_4}(red). The rate increases sub-linearly up to $N=3$, and then slowly decreases as more resonators are added to the sequence. 
\\
At first glance this result is puzzling. 
Indeed, one might argue that even in the case of independent emitters, 
the probability that at least one ring fires should scale with $N$. However, here we have to consider that the pump and the signal/idler photons are attenuated by each drop event, thus the coincidence rate 
reaches a maximum, then 
decreases when the losses prevail. 
Still, it is not clear whether this result is associated with any cooperative behaviour, for a similar trend could be
also expected for incoherent emission of each ring individually. To test this hypothesis, each source $S_j$ is pumped individually with an input power of $P_j [\textrm{dB}] =P_1[\textrm{dB}]-(j-1)T_d[\textrm{dB}]$, where $T_d$ is the on-resonance drop transmittance. The scaling of the pump power is made necessary to take into account the losses after each drop event. We then measured the coincidence rates $C_j$, and calculated the expected incoherent rate $R(N)$ at the end of the sequence as $R(N)=\sum_j^N T_{jN}C_j$, where $T_{jN}$ is the effective transmittance of the pair from source $j$ to $N$ (see calculation in Appendix D). The quantity $R(N)/R(1)$ is reported in Fig.\ref{Fig_4} (black).  \textcolor{black}{The incoherent sum is approximately constant up to $N=3$, and then drops as the size of the array is increased. This clearly disagrees with the result for the collective excitation of the emitters (red line in Fig.\ref{Fig_4}).} 
and demonstrates the presence of cooperative effects in the array that enhance the pair generation probability. 

Following \textcolor{black}{Onodera et al.} \cite{onodera2016parametric}, cooperativity \textcolor{black}{is associated with} 
the impossibility \textcolor{black}{of distinguishing} 
which ring emitted the pair that is detected at the output of the sequence, from which quantum interference arises. In the ideal 
\textcolor{black}{scenario}  where all the sources are 
\textcolor{black}{identical} and fully coherent, the pair generation probability is predicted to be enhanced by a  factor $N^2$ 
\textcolor{black}{over that of} a single emitter. It is worth to stress that this result does not trivially follows from the intrinsic scaling of SFWM with the system size $\mathscr{L}$. Even in a bare waveguide, the SFWM intensity grows as $\mathscr{L}^2$ only when photon pairs are spectrally filtered with a bandwidth that is much smaller than the one determined by dispersion, while in the unfiltered case the scaling is $\mathscr{L}^{\frac{3}{2}}$ \cite{helt2012does}. 
The $N^2$ dependence is rather a consequence of quantum interference. To this end, it is instructive to consider the analogy with the single photon super-radiance in atomic clouds \cite{scully2009super}, and view super SFWM as an enhanced collective decay process. In this  picture, the pump beam interacts with all the rings and probabilistically generates a photon pair in one of them, i.e., it creates an excitation that is symmetrically stored in the array of emitters. In the atomic counterpart, this state corresponds to the case where only one atom is fully excited by the absorption of one photon, but we do not know which one. Then, each resonator can "spontaneously emit" the pair into the bus waveguide, in the very same way a two level atom can decay into a radiation mode via its electric dipole coupling to the vacuum of the electromagnetic field (indeed, the two processes are described by formally equivalent Hamiltonians \cite{scully2006directed,onodera2016parametric,helt2012does}). The $N$ transition amplitudes 
associated to those events where pairs generated in source $j$ are scattered towards port $D_N$ are all coherent and with a negligible relative phase (see Appendix A for derivation). Constructive interference \textcolor{black}{would} 
then \textcolor{black}{be} expected to occur among these transition amplitudes, enhancing the decay process by $N$ times compared to the case of an incoherent emission, and completing the analogy with the single photon super-radiance \cite{scully2006directed,scully2009super}. 
\textcolor{black}{However,}  in the real scenario the marginal spectra of photons generated by different rings are partially distinguishable. Indeed, pairs emitted by ring $j$ are dropped and filtered $N-j$ times before they reach the end of the sequence, a fact that progressively narrows their spectral linewidth and decreases the effective brightness of the source. 

To bring our system closer to the ideal conditions, we considered FWM and suppressed the filtering effect by stimulating the generation of signal photons by resonantly injecting a CW idler seed. As shown in Fig.\ref{Fig_1}(c), the seed is combined to the pump laser using a $50/50$ beamsplitter before entering at the input of the chip. Energy conservation forces the signal beam to have a bandwidth comparable to that of the pump and the idler laser (\textcolor{black}{a} few MHz), which is much smaller than that of the resonators, therefore eliminating the filtering issue. The stimulated signal is detected at the output of the sequence using a monochromator coupled to a CCD camera, and its intensity is shown in Fig.\ref{Fig_4} (blue) as a function of the size of the array. The growth of this curve is super-linear up to $N=3$, a genuine manifestation of  coherent emission of the resonators. 

The larger scaling of stimulated FWM with respect to the spontaneous process comes from two contributions. First, the narrowband emission eliminates the effect of spectral filtering. Second, the stimulated process intrinsically scales better with $N$ than the spontaneous case, even in absence of spectral filtering\textcolor{black}{; this} 
is due to the different way the losses are impacting the rate of the singles and the coincidences. The first 
\textcolor{black}{contribution dominates} 
in our 
\textcolor{black}{experiment} (see Appendix C for demonstration). 
Still, lossess prevent even for the stimulated process to scale with $N^2$. 

\begin{figure}[h!]
\centering\includegraphics[scale = 0.61]{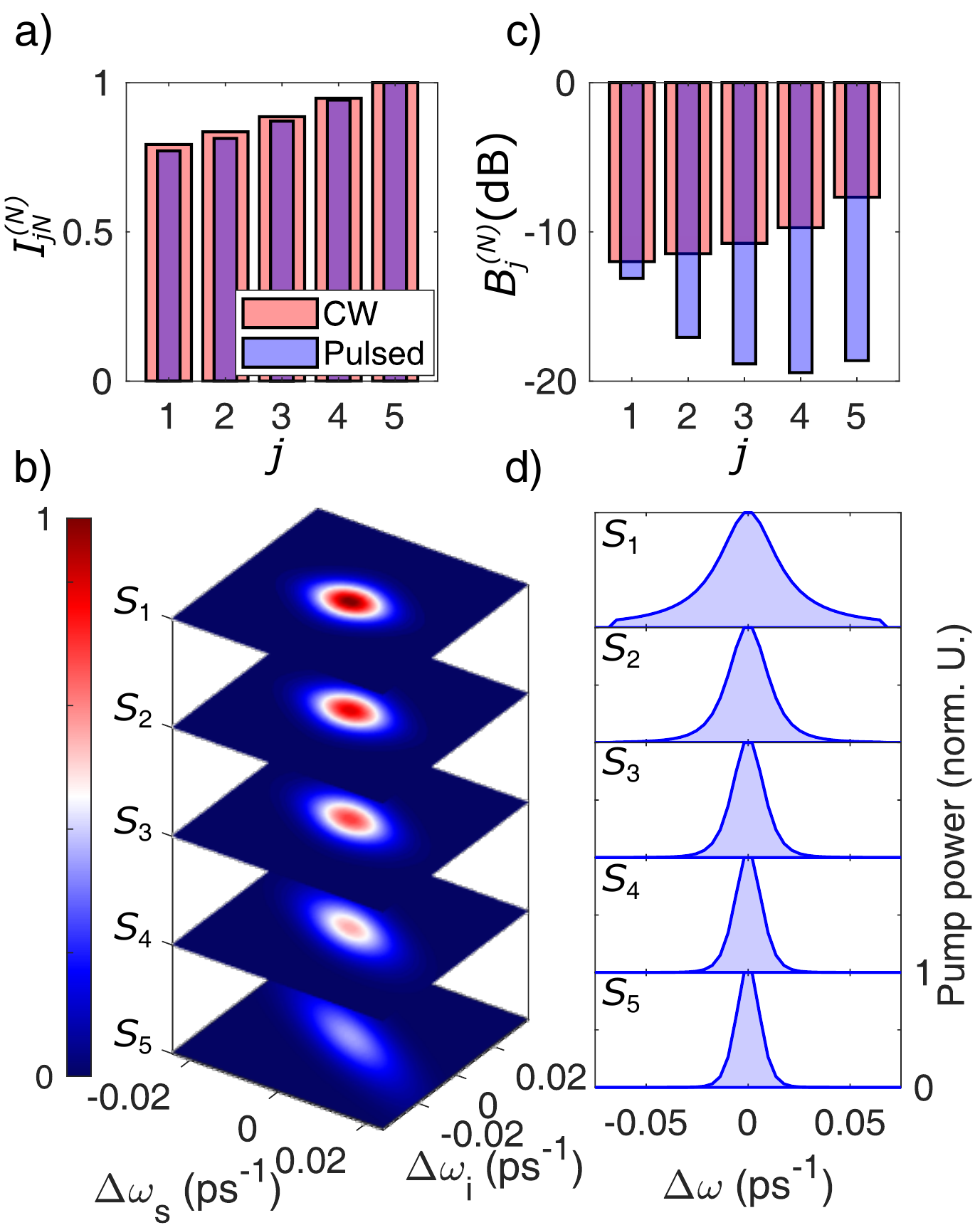}
\caption{(a) Indistinguishability $I_{jN}$ between source $j$ and source $N$ in a sequence of $N=5$ resonators. (b) Modulus square of the JSA of the different sources in a sequence of $N=5$ resonators, as seen from the output port $D_5$. The simulation uses a pulsed pump at the input (panel (d)). (c) Relative (compared to a single resonator) brightness $B_{j}^{(N)}$ of source $S_j$ in a sequence of $N=5$ resonators. (d)  Pump spectrum at the input of each source $S_j$ along a sequence of $N=5$ resonators. \label{Fig_5}}
\end{figure}

We \textcolor{black}{can identify} 
the on-resonance drop transmission $T_d$  as the ultimate factor limiting the full cooperativity of the array. In Appendix C, it is proved that when losses are considered ($T_d<1$), the signal intensity scales as $T_d^{N-1}((1-T_d^N)/(1-T_d))^2$. This relation is used to fit the experimental data for stimulated FWM, showing the best agreement (dashed blue line in Fig.\ref{Fig_4}) when $T_{d,\textrm{fit}}=0.75\,(-1.25\,\textrm{dB})$, which is close to the measured value. On the contrary, a model based on incoherent emitters predicts a scaling law $T_d^{N-1}(1-T_d^{2N})/(1-T_d^2)$, which does not reproduce the experimental data for any meaningful value of $T_d$. To further highlight the critical role of losses on the cooperativity in the spontaneous process, we repeated the experiment using a broadband pulsed pump laser. The (filtered) bandwidth ($\sim 80\,\textrm{pm}$) is intentionally set to be approximately twice the one of the resonators to introduce additional filtering loss at each drop event. As shown in Fig.\ref{Fig_1}(c), this is obtained by  filtering a broad-band (few nm) femtosecond laser with a Fiber Bragg Grating and a circulator. 
As shown in Fig.\ref{Fig_4} (green), the measured coincidence rates are now much smaller than that of the CW case. To consolidate these observations, we modeled the pair generation process using the method of asymptotic fields described 
\textcolor{black}{earlier}  \cite{liscidini2012asymptotic}. We leave the detailed derivation to Appendix A, 
and report here the final expression for the normalized rate $R(N)/R(1)$, which is given by
\begin{equation}
    \frac{R(N)}{R(1)} = \sum_{j,k=1}^{N}\sqrt{B_j^{(N)}B_k^{(N)}} I_{jk}^{(N)},
    \label{eq:1}
\end{equation}
\noindent where $B^{(N)}_j$ is the relative (compared to a single resonator) brightness of source $j$ as seen from the output of the sequence of $N$ rings. 
The quantity $I^{(N)}_{jk}$ is the indistinguishability between the 
\textcolor{black}{joint spectral amplitude} (JSA) of photon pairs emitted by source $j$ and $k$, as seen from the output port $D_N$ \cite{borghi2020phase}. In Fig.\ref{Fig_5}(a) we plot $I^{(N)}_{jk}$ for $k=N$ and $N=5$, 
\textcolor{black}{for} both 
CW and the pulsed 
\textcolor{black}{pumping.} The indistinguishability between source $S_j$ and source $S_5$ monotonically increases with $j$. This is due to the spectral filtering on the pairs, which is manifested by the progressive narrowing of the JSA as we proceed backwards from the end of the sequence (Fig.\ref{Fig_5}(b)). As shown in Fig.\ref{Fig_5}(c), the relative brightness \textcolor{black}{$B_j^{(N)}$} is greatly reduced 
\textcolor{black}{as $j$ increases for pulsed pumping, while it increases as $j$ increases for CW pumping.} 
This is caused by the spectral filtering on the broadband pump, shown in Fig.\ref{Fig_5}(d), which acts as an excess loss. 
We used Eq.(\ref{eq:1}) to simulate the expected rates 
\textcolor{black}{for} both 
CW and pulsed \textcolor{black}{pumping}, 
leaving only the drop transmittance $T_d$ as a free parameter. We found the best agreement when $T_{d,\textrm{fit}}=0.8$ in the CW case, and $T_{d,\textrm{fit}} = 0.79$ in the pulsed case. 
\\
\noindent It is now natural to ask if a super SFWM regime, i.e., a rate of emission which scales as $\frac{R(N)}{R(1)}>N^b$ with $1\leq b\leq 2$, could be reached also in the spontaneous case in lossy integrated systems. By looking at Eq.(\ref{eq:1}), this is  possible if $\sqrt{B_j^{(N)}B_k^{(N)}} I_{jk}^{(N)}>\frac{1}{N}$, a requirement that can be met by lowering the drop and the filtering losses. As an example, the first can be reduced by increasing the coupling $\kappa^2$ between the resonator and the waveguide, as the drop transmittance scales as $T_d = \left ( 1-\frac{1}{2(1+\xi)}\right )^2$, where $\xi=\frac{Q_i}{Q_e}$, is the ratio between the intrinsic ($Q_i$) and extrinsic ($Q_e\propto \frac{1}{\kappa^2}$) quality factors. It can be proved that in the ideal (lossless scenario where $T_d=1$, the presence of spectral filtering still limits the scaling to $N^{\frac{3}{2}}$ (see Appendix B for demonstration).
This issue could be mitigated by gradually increasing the linewidth of signal/idler resonances along the sequence, for example through an apodization of the quality factors. 
Improving the super SFWM rate closer to the theoretical limit of $N^2$ could open several perspectives of both technological and fundamental relevance. First, the brightness of the array would be improved without increasing the input pump power, which is appealing for the realization of ultra-bright photon pair sources by preventing the saturation of the pump power induced by nonlinear absorption in silicon \cite{engin2013photon,husko2013multi}. Second,  we foresee that by post-selecting a higher number of sources which fire simultaneously, we could use the integrated photonic platform to explore other cooperative regimes of many body physics with high process fidelity. Finally, our results may foster future works exploring the coherent combination of light from several microrings to construct sources based on cooperative super SFWM. \\
\section{Conclusions}
In conclusion, we have  provided experimental evidence of cooperative emission of photon pairs from an array of microresonators on a silicon photonic chip, bringing the original theoretical study from Onodera et. al \cite{onodera2016parametric} from theory to practice.
We identified losses and indistinguishability as the main factors limiting the super SFWM scaling. These could be greatly mitigated by improving the device design. 
Our work shows that the integrated platform is a reliable alternative to 
those based on cold matter systems, offering a robust control on the state evolution, design possibilities and intrinsic scalability. The state associated to a single pair emission is in itself linked to intriguing physics, as single photon superradiance \cite{tighineanu2016single,scully2009super} or directional spontaneous emission \cite{scully2006directed}. The natural extension to multiple pair emission opens the possibility of using silicon photonic devices for the emulation of many body atomic systems of increased complexity. 
\\
\acknowledgements{
This work has been supported by Ministero dell’Istruzione,
dell’ Università e della Ricerca [MIUR grant Dipartimenti
di Eccellenza 2018-2022 (F11I18000680001)]. J. E. S.
acknowledges support from the Natural Sciences and
Engineering Research Council of Canada. The device has
been designed using the open source Nazca design\texttrademark \,framework.}
\section*{Appendix A: Model of pair generation from the array of resonators}
\begin{figure}[t!]
\centering\includegraphics[scale = 0.45]{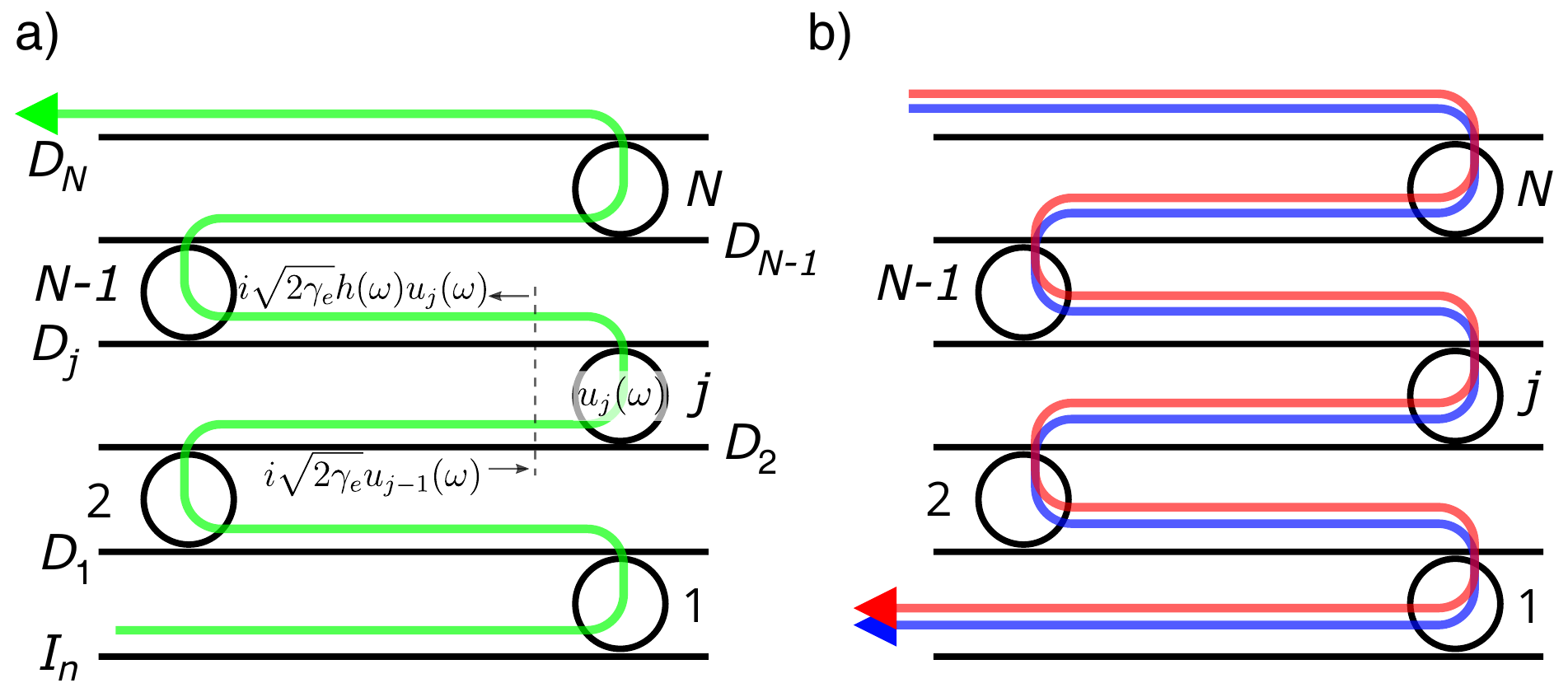}
\caption{(a) A sketch of the asymptotic input field of the pump. The energy amplitude inside the resonators is indicated with $u_j$. The formulas close to the arrows are the field amplitudes entering and leaving the nonlinear interaction region of ring $j$. The green arrow follows the energy flow of the pump from the input to the output of the sequence. (b) Sketch of the time reversal of the asymptotic output field of the signal (red) and idler (blue) photon. \label{Fig_6}}
\end {figure}

\noindent We used the formalism of asymptotic fields to model SFWM from the array of resonators \cite{liscidini2012asymptotic,yang2008spontaneous}. We expanded the electric displacement field $\mathbf{D}(\mathbf{r})$ in terms of asymptotic input fields $\mathbf{D}^{\textrm{in}}(\mathbf{r})$ for the pump, while we used asymptotic output fields $\mathbf{D}^{\textrm{out}}$ for the signal/Idler,
\begin{multline}
    \mathbf{D}(\mathbf{r})=\int_{k\in \{k_{p_0}\}}\sqrt{\frac{\hbar\omega_k}{2}} \mathbf{d}_k(\mathbf{r}) a_k dk \\ + \int_{k\in \{k_{s_0},k_{i_0}\}}\sqrt{\frac{\hbar\omega_k}{2}} \mathbf{d}_k(\mathbf{r}) b_k dk +  \textrm{h.c.}, \label{eq:S4}
\end{multline}
where $a_k$ and $b_k$ are the photon lowering operators at wavevector $k$ and frequency $\omega_k$ for the pump and signal/idler photons respectively. The functions $\mathbf{d}_k(\mathbf{r})$ represent the normalized spatial profiles of the asymptotic field in the structure. The integrals in Eq.(\ref{eq:S4}) are performed in the neighbour of $\{k_{s_0},k_{p_0},k_{i_0}\}$, i.e., the wavevectors of the three, not overlapping frequency intervals around the signal, the pump and the idler resonance. With reference to Fig.\ref{Fig_6}(a), the field $\mathbf{D}^{\textrm{in}}$ describes a wave entering at the input port of the array which is routed to the drop of the last ring $N$ of the sequence (in the following, all the resonators are assumed to be identical and spectrally overlapped). The energy amplitude inside  ring $j$ is denoted as $u_j$. Similarly, the time reversal $(\mathbf{D}^{\textrm{out}})^*$ of the field $(\mathbf{D}^{\textrm{out}})$, shown in Fig.\ref{Fig_6}(b), corresponds to a wave entering from port $D_N$ and leaving the array from the input. 
The reason why we consider $(\mathbf{D}^{\textrm{out}})^*$ and not $(\mathbf{D}^{\textrm{out}})$ will be clarified later. We stress that the form of the asymptotic fields shown in Fig.\ref{Fig_6} approximates the real scenario, where photons can also be scattered into the Through ports or away from the chip. In principle, their contribution has to be included  to ensure that the system 
\textcolor{black}{Hamiltonian} generates a unitary evolution of the states. However, since we are only interested into those cases where signal/idler pairs leave together ports $D_j$, we can obtain the same result of the full calculation by using a transfer function \textcolor{black}{relating $u_{j+1}$ to $u_j$} 
that is not energy conserving. The approach is analogous to the one described 
\textcolor{black}{earlier} \cite{banic2021modeling}, where the effect of scattering losses in pair generation is accounted \textcolor{black}{for} by introducing an effective absorption coefficient in the asymptotic fields.  When the expression in Eq.(\ref{eq:S4}) is inserted into the 
\textcolor{black}{Hamiltonian} $H$ of the system, we get $H=H_0+V^{(N)}$, where $H_0$ governs the free evolution of the asymptotic fields, while $V^{(N)}$ describes the third order nonlinear interaction in the array of $N$ rings, which is responsible of SFWM \cite{yang2008spontaneous,helt2012does}. The  operator $V^{(N)}$ 
\textcolor{black}{can be written as}
\begin{equation}
    V^{(N)} = - \int d\mathbf{k} S^{(N)}(k_1,k_2,k_3,k_4)b_{k_1}^{\dagger}b_{k_2}^{\dagger}a_{k_3}a_{k_4} + \textrm{h.c.},
\end{equation}
where $d\mathbf{k}=dk_1dk_2dk_3dk_4$ and
\begin{equation}
    S^{(N)} \sim \frac{3(\Gamma_3 \hbar\omega_{p_0})^2}{32\epsilon_0} \int J^{(N)}(k_1,k_2,k_3,k_4) d\mathbf{k}\label{eq:S6},
\end{equation}
\begin{equation}
    J^{(N)} = \int (d^{\textrm{out}}_{k_1}(\mathbf{r}))^*(d^{\textrm{out}}_{k_2}(\mathbf{r}))^* d^{\textrm{in}}_{k_3}(\mathbf{r})d^{\textrm{in}}_{k_4}(\mathbf{r}) d\mathbf{r} \label{eq:S7}.
\end{equation}
\textcolor{black}{For} simplicity we used a scalar form of $\mathbf{d}_k(\mathbf{r})$, keeping only to the contribution of the dominant component of the field. The quantity $\Gamma_3$ is a constant related to the nonlinear effective index of the material, the effective area of the waveguide and to the pump power \cite{helt2012super}.
Due to the form of $V^{(N)}$, the asymptotic output fields enter with a complex conjugate in Eq.(\ref{eq:S7}), which motivates the choice to consider the time reversal of $\mathbf{D}^{\textrm{out}}$ in Fig.\ref{Fig_6}(b). In principle, the volume integral in Eq.(\ref{eq:S7}) should be evaluated both in the rings and in the waveguide sections connecting them. In practice, since the field intensity is  more intense inside the resonators, we restrict the calculation to that region. In the case of the pump, we derive the field intensity inside ring $j$ starting from the energy amplitude $u_j$ predicted by TCMT 
\cite{borghi2020phase}; 
\textcolor{black}{it is given} by
\begin{equation}
    u_j(\omega) = h(\omega)s_j = -\frac{i\sqrt{2\gamma_e}}{i(\omega-\omega_0)+\gamma_{\textrm{tot}}}s_{j} \label{eq:S8},
\end{equation}
where $s_j$ is the incoming wave at the input of ring $j$ and $h(\omega)$ can be viewed as a transfer function. By looking at Fig.\ref{Fig_6}(a), the input $s_j$ is given in turn by $s_j=i\sqrt{2\gamma_e}e^{ik(\omega)L}u_{j-1}$, where $L$ is the separation between the resonators, which inserted into Eq.(\ref{eq:S8}) gives $u_j(\omega)=i\sqrt{2\gamma_e}h(\omega)e^{ik(\omega)L}u_{j-1}$. It is worth \textcolor{black}{noting} 
that $2\gamma_e\int |h(\omega)|^2d\omega\leq1$, which implies that some photons are lost during the propagation from $D_{j-1}$ to $D_j$. These can 
either \textcolor{black}{exit via} 
the Through port or 
\textcolor{black}{be scattered off the chip.} 
We can use this recursive relation $j-1$ times to relate $u_j$ to $s_1=1$, i.e., \textcolor{black}{to }the excitation amplitude at the input,
\begin{multline}
    u_j(\omega) = (i\sqrt{2\gamma_e})^{j-1}e^{i(j-1)k(\omega)L}h(\omega)^j \\ \textrm{(pump)}.\label{eq:S9}
\end{multline}
Similarly, we can construct an equivalent relation for the asymptotic output fields \textcolor{black}{of the signal and idler} inside the rings. 
This time, the number of resonators 
\textcolor{black}{preceding} $j$ is $N-j$, hence
\begin{multline}
    u_j(\omega) = (i\sqrt{2\gamma_e})^{N-j}e^{i(N-j)k(\omega)L}h(\omega)^{N-j+1} \\ \textrm{(signal,\,idler)},\label{eq:S10}
\end{multline}
where $h(\omega)$ has to be considered centered at the 
resonance frequency of the signal \textcolor{black}{or} 
idler. From now on, \textcolor{black}{we} 
label the transfer function of the three waves as $h_p,\,h_s$ and $h_i$. The displacement field inside ring $j$ is related to $u_j$ \cite{onodera2016parametric} by 
\begin{equation}
    d^{(j)}_{k}(\mathbf{r}) = \frac{u_j(\omega(k))}{\sqrt{2\pi\tau_{\textrm{rt}}}}e^{ik\xi}\mathscr{D}_k(\mathbf{r_t}), \label{eq:S11}
\end{equation}
where $\tau_{\textrm{rt}}$ is the cavity round trip time, $\xi$ is the azimuthal coordinate along the ring, $\mathbf{r_t}$ are the transverse coordinates spanning the waveguide cross section, and $\mathscr{D}_k$ is the mode profile. Inserting Eq.(\ref{eq:S11}) into Eq.(\ref{eq:S7}), and using the expressions in Eq.(\ref{eq:S8}-\ref{eq:S9}), we get $ J^{(N)}\propto \sum_{q=1}^N j_q^{(N)} e^{iq\Delta kL}$ where $j_q$ is given by
\begin{multline}
    j_q^{(N)} = (-2\sqrt{\gamma_{e,s}\gamma_{e,i}})^{N-q}(-2\gamma_{e,p})^{2(q-1)} \\  h^q_p(\omega_{k_3})h^q_p(\omega_{k_4})h^{N-q+1}_s(\omega_{k_1})h^{N-q+1}_i(\omega_{k_2}), \label{eq:S12}
\end{multline}
where all the terms that do not depend on $j$ or $N$ have been factored out to simplify the expression, and we have defined $\Delta k=k_3+k_4-k_1-k_2$.  The quantity $J^{(N)}$ enters in the expression of the \textcolor{black}{joint spectral amplitude} 
\cite{onodera2016parametric,helt2012super} as
\begin{multline}
    \phi^{(N)}(\omega_1,\omega_2) = \frac{\mathscr{K}}{\beta_N}\int \phi_p(\omega)\phi_p(\omega_1+\omega_2-\omega) \\ J^{(N)}(\omega_1,\omega_2,\omega,\omega_1+\omega_2-\omega)d\omega, \label{eq:S13}
\end{multline}
where $\phi_p$ is the normalized spectral amplitude of the pump, $\mathscr{K}$ is a constant, and $|\beta_N|^2$ is the pair generation probability from the array of $N$ rings.
By substituting Eq.(\ref{eq:S12}) into the expression of $J^{(N)}$ in Eq.(\ref{eq:S13}) we get \mbox{$\phi^{(N)}(\omega_1,\omega_2) = \frac{\mathscr{K}}{\beta_N}\sum_{q=1}^N \varphi^{(N)}_q(\omega_1,\omega_2)$}, where
\begin{multline}
    \varphi^{(N)}_q(\omega_1,\omega_2) = \int \phi_p(\omega)\phi_p(\omega_1+\omega_2-\omega) \\ {j^{(N)}}_q(\omega_1,\omega_2,\omega,\omega_1+\omega_2-\omega) e^{iq\Delta k L}d\omega, \label{eq:S14}
\end{multline}
and from the normalization condition, $\int |\phi^{(N)}(\omega_1,\omega_2)|^2 d\omega_1 d\omega_2 = 1$, \textcolor{black}{we have}
\begin{equation}
    |\beta_N|^2=\sum_{\{j,k\}=1}^N \sqrt{{B^{\prime (N)}}_j {B^{\prime(N)}}_k}{I^{(N)}}_{jk}e^{i(j-k)\bar{\Delta k}L}, \label{eq:S15}
\end{equation}
with the definitions
\begin{align}
    B^{\prime(N)}_j = & {\mathscr{K}}^2\int {{|\varphi^{(N)}}_j(\omega_1,\omega_2)|}^2 d\omega_1 d\omega_2, \label{eq:S16} \\
    {I^{(N)}}_{jk} = & \frac{\int {\varphi^{(N)}}_j(\omega_1,\omega_2){\varphi^{(N)}_k}^*(\omega_1,\omega_2)d\omega_1 d\omega_2}{\sqrt{{B^{\prime (N)}}_j {B^{\prime (N)}}_k}}. \label{eq:S17}
\end{align}
In writing Eq.(\ref{eq:S15}), we replaced \textcolor{black}{the function} $\Delta k=k(\omega)+k(\omega_1+\omega_2-\omega)-k(\omega_1)-k(\omega_2)$ \textcolor{black}{by} 
the 
value $\bar{\Delta k}=2k_{p_0}-k_{s_0}-k_{i_0}$, since $\Delta kL$ is considered to vary slowly over the wavevector range where the integral in Eq.(\ref{eq:S14}) has not vanishing contributions. This condition holds as long as $NL \ll \frac{\pi}{\Delta k}=L_{\textrm{coh}}$, where $L_{\textrm{coh}}$ is the coherence length of the FWM interaction. 
In our 
\textcolor{black}{experiment,} this quantity is of the order of few cm for signal/idler pairs that are generated within $\sim 20$ nm of spectral distance from the pump, which justifies the approximation. We can interpret $B^{\prime (N)}_j$ as the pair generation probability of source $j$ in the array, and $I^{(N)}_{jk}$ as the indistinguishability between the JSA of source $j$ and source $k$ (as seen from the output port $D_N$). Terms with $j\neq k$ represent the interference between the amplitude probabilities of generating a pair  in ring $j$ and $k$, and have a relative phase $(j-k)\bar{\Delta k}L=\pi(j-k)\frac{L}{L_{\textrm{coh}}}$. Since $NL\ll L_{\textrm{coh}}$, we have that $(j-k)\Delta k L \ll 1$, so all these terms interfere constructively.  \\

\noindent The coincidence rate $R(N)$ between the signal and the idler photons at the output of the  sequence is proportional to $|\beta_N|^2$, hence the normalized rate $\frac{R(N)}{R(1)}=\left |\frac{\beta_N}{\beta_1} \right |^2$ 
\textcolor{black}{can be} written as
\begin{equation}
   \frac{R_N}{R_1}=\sum_{\{j,k\}=1}^N \sqrt{B^{(N)}_jB^{(N)}_k}I^{(N)}_{jk}, \label{eq:S18}
\end{equation}
where $B^{(N)}_j=\frac{B^{\prime (N)}_j}{B^{\prime (1)}}$ is the relative (compared to a single resonator) brightness of source $j$ as seen from the output of the sequence of $N$ rings. This expression coincides with Eq.(\ref{eq:1}) of the main text. The quantities $B^{(N)}_j$ and $I^{(N)}_{jk}$ have been computed through numerical simulations using Eq.(\ref{eq:S16},\ref{eq:S17}), and are shown in the Supplemental Fig.3 \cite{suppl} for both CW and pulsed excitation. Using these values, we fit the scaling laws shown in Fig.\ref{Fig_4}. In all the simulations, the values of the loaded quality factors $Q_{\textrm{tot}}=\frac{\omega}{2\gamma_{\textrm{tot}}}$ and the shape of the pump spectra are taken from the experiment, as detailed in the Supplemental Material \cite{suppl}. We left the drop transmittance $T_d$ as a free parameter, from which the intrinsic quality factor  $Q_i=\frac{\omega}{2\gamma_i}$ can be computed as $Q_i=Q_{\textrm{tot}}/(1-\sqrt{T_d})$. The values of $T_{d,\textrm{fit}}$ that minimize the least square error with the experimental data are reported in Table \ref{Tab_2}.
\begin{table}[h!]
\begin{centering}
\begin{tabular}{c|c}
Process & $T_{d,\textrm{fit}}$  \\
\hline 
Stimulated FWM & 0.75\\ 
\hline 
Spontaneous FWM (CW) & 0.8 \\
\hline 
Spontaneous FWM (pulsed) & 0.79 
\end{tabular}
\end{centering}
\caption{Values of the drop transmittance $T_{d,\textrm{fit}}$ which minimize the least square error with the experimental data shown in Fig.\ref{Fig_4}. \label{Tab_2}}
\end{table}

\section*{Appendix B: Asymptotic scaling for long sequences of resonators}
\noindent In this section, we derive the asymptotic behaviour of the pair generation rate for a large number $N$ of resonators in the CW regime. This is formally a generalization to a chain of lossy add-drop resonators of the ideal scenario described 
\textcolor{black}{earlier} \cite{onodera2016parametric}, where a lossless sequence of all-pass resonator 
\textcolor{black}{was} considered. 
\textcolor{black} {The} authors found that under CW pumping, the pair generation probability scales as $N^2$ with the number of resonators. Here we 
show that without apodization of the resonance linewidth (see main text), the filtering loss limit\textcolor{black}{s} the asymptotic scaling of the spontaneous rate to $N^{\frac{3}{2}}$, i.e., an intermediate value lying between the fully incoherent case ($\propto N$) and the ideal case without spectral filtering \cite{onodera2016parametric}. We start by noting that, following Eq.(\ref{eq:S8}) and the discussions above, we can relate the internal energy $u_j$ to the excitation $s_{j-1}$ as
\begin{equation}
    u_j(\omega)=t(\omega)e^{ik(\omega)L}h(\omega)s_{j-1} \label{eq:SS1},
\end{equation}
where we used $s_j=t(\omega)e^{ik(\omega)L}s_{j-1}$ and \textcolor{black}{have} introduced the drop transmittance $t(\omega)$, which is given by
\begin{equation}
    t(\omega)=\frac{2\gamma_e}{i(\omega-\omega_0)+\gamma_{\textrm{tot}}}. \label{eq:SS2}
\end{equation}
By iterating Eq.(\ref{eq:SS1}) $j-1$ times, we obtain \mbox{$u_j(\omega)=t^{j-1}(\omega)h(\omega)e^{i(j-1)k(\omega)L}s_1$} for the pump internal energy, and $u_j(\omega) = t^{N-j}(\omega)h(\omega)e^{i(N-j)k(\omega)L}s_1$ for that of the signal/idler. When this expression is plugged into Eq.(\ref{eq:S7}), through the use of Eq.(\ref{eq:S11}) and a little bit of algebra we have
\begin{equation}
J^{(N)}\propto T_d^{N}\sum_{i=1} ^{N}  L(\omega^{\prime})^{N-i+1} =T_d^{N}\sum_{i=1} ^{N}  L^{i}(\omega^{\prime}),
\label{eq:SS3}
\end{equation}
where $T_d=|t(\omega_{0p})|^2=|t(\omega_{0s})|^2=|t(\omega_{0i})|^2$ equals the on resonance transmitted intensity of the resonator at each of the pump, idler, and signal resonant frequencies, while 
\begin{equation}
L(\omega^{\prime})= \frac{ \gamma_{\textrm{tot}}^2}{{\omega^{\prime}}^2+ \gamma_{\textrm{tot}}^2},
\label{eq:Lorentzian}
\end{equation}
is a \textcolor{black}{L}orentzian function. In Eq.(\ref{eq:Lorentzian}), we \textcolor{black} {write} 
\mbox{$\omega^{\prime}=\omega-\omega_{0s}$} 
\textcolor{black}{as} the frequency shift from the signal resonance at $\omega_{0s}$. The modulus square of $J^{(N)}$ is then given by
\begin{multline}
\left| J(\omega^{\prime})^{(N)}\right|^2 \propto  \left|T_{d}\right|^{2N} \left(\sum_{i=1} ^{N}  L^{i}(\omega^{\prime})\right)^2 = \\ \left|T_{d}\right|^{2N}
\left(\sum_{i=1} ^{N} i L^{i+1}(\omega^{\prime})+ \sum_{i=1} ^{N-1} i L^{2N-i+1}(\omega^{\prime})\right).
\label{eq:J2  approx}
\end{multline}
The pair generation probability $|\beta_N|^2$ is proportional to $\int |J^{(N)}(\omega^{\prime}|^2d\omega^{\prime}$, which demands to calculate the integral of powers of a \textcolor{black}{L}orentzian function. This can be achieved with the aid of the $T_d$ function,
\begin{equation}
\int_{-\infty} ^{\infty} {  L^i(\omega^{\prime})d\omega^{\prime}}=\sqrt{\pi}\gamma_{\textrm{tot}}\frac{\Gamma_d(i-0.5)}{\Gamma_d(i)} \sim \gamma_{\textrm{tot}}\sqrt{\frac{\pi}{i}},
\label{eq:Lorentzian int}
\end{equation}
where in the last step we used the asymptotic formula for $x \to \infty$ for the $\Gamma_d$ function $\Gamma_d(x+\alpha)=\Gamma_d(x)x^\alpha$.
Using Eq.(\ref{eq:Lorentzian int}), the pair generation rate is
\begin{equation}
\left| \beta_N \right|^2 \propto \left|T_{d}\right|^{2N}  
\times \left(\sum_{i=1} ^{N} \frac{i}{\sqrt{i+1}} + \sum_{i=1} ^{N-1}\frac{i}{\sqrt{2N-i+1}}\right).
\label{eq:Pair generation approx}
\end{equation}
The first summation can be approximated with a similar integral,
\begin{equation}
\int_{0}^{N}{\frac{x}{\sqrt{x+1}}dx}= \frac{2}{3}\sqrt{(N + 1)}(N - 2) + 4/3\sim N^{3/2}.
\label{eq:sum of sqrts}
\end{equation}
The second summation scales as the former for large values of $N$, hence the pair generation probability asymptotically scales as $\left| \beta_N \right|^2 \propto \left|T_{d}\right|^{2N} N^{\frac{3}{2}}$ with the number $N$ of resonators. In the ideal scenario where $T_{d}=1$, which corresponds to the limit of vanishing propagation losses, the spectral filtering ultimately limits the   scaling to $N^{\frac{3}{2}}$. In a typical experiment one has $N\sim\mathscr{O}(10)$, and for such value the asymptotic expressions in Eq.(\ref{eq:Lorentzian int}) and in Eq.(\ref{eq:sum of sqrts})  already appliy to a good approximation ($\Gamma(N+0.5)\approx0.99\times \Gamma(N)N^{0.5}$ and $\frac{2}{3}\sqrt{(N + 1)}(N - 2) + 4/3\approx 0.6N^{\frac{3}{2}}$ for $N=10$, the latter being compared to the asymptotic behaviour $\sim \frac{2}{3}N^{\frac{3}{2}}\approx0.66N^{\frac{3}{2}}$).

\section*{Appendix C: Impact of the array losses on the intensity of stimulated and spontaneous four-wave mixing}
\noindent Here, we give a simplified treatment describing the  impact of loss in the scaling of the intensity of stimulated and spontaneous FWM from the array of resonators. We consider a CW excitation and neglect any effect related to spectral filtering and distinguishability from the different sources. With reference to Fig.\ref{Fig_1}(b), the stimulated field amplitude at the output port $D_N$ is given by $\sum_{j=1}^N A_j$, where $A_j$ is the contribution of ring $j$ to the total field reaching port $D_N$. By denoting with $P_p$ and $P_i$ the powers of the pump and of the stimulating  laser at the input of the sequence, and with $T_d$ the drop transmittance (assumed for simplicity to be equal at both the pump, the signal and the idler wavelengths), we have that the stimulated field immediately \emph{after} the port $D_j$ is $A^{\prime}_j=\kappa_1 (T_d^{\frac{j-1}{2}}P_i)(T_d^{(j-1)}P_p^2)$, where $\kappa_1$ incorporates all the factors that do not depend on $j$ or $N$. The field $A^{\prime}_j$ is dropped $N-j$ times before it reaches port $D_N$, hence $A_j=T_d^{\frac{N-j}{2}}A^{\prime}_j=T_d^{\frac{N-3}{2}}T_d^{j}\kappa_1 P_p^2 P_i$. The normalized power $\xi_{\textrm{stim}} = \frac{P_i(N)}{P_i(1)}$ is then given by
\begin{equation}
\xi_{\textrm{stim}}=\left | \sum_{j=1}^NA_j \right |^2=T_d^{N-1}\left ( \frac{1-T_d^N}{1-T_d} \right )^2,\label{eq:S2}
\end{equation}
where we used the result from the geometric series $\sum_{j=m}^{N}\textcolor{black}{x^j}=(x^m-x^{N+1})/(1-x)$. It is worth 
\textcolor{black}{noting} that $\frac{P_i(N)}{P_i(1)}\rightarrow N^2$ as $T_d \rightarrow 1$.  We now compare the scaling of stimulated FWM to that of the spontaneous case. As stated before, we neglect the effect of spectral filtering on the pairs imparted by the cascade of resonators in the sequence. This can be experimentally realized by placing a bandpass filter, with a bandwidth much smaller than that of the resonators, before the detection of the signal and the idler photons. Under these conditions, the amplitude probability of generating a pair at port $D_j$ simplifies to $A^{\prime}_j=\kappa_2P_p^2 T_d^{j-1}$, which after $N-j$ drop events becomes $A^{\prime}_jT_d^{N-j}=\kappa_2 P_p^2 T_d^{N-1}$. Note that, in contrast with stimulated FWM, the inclusion of both 
signal and the idler losses  
\textcolor{black}{leads to an} $A_j$ 
\textcolor{black}{that does} not depend on $j$. The normalized coincidence rate $\xi_{\textrm{spont}}=\frac{R(N)}{R(1)}$ then scales as:
\begin{equation}
\xi_{\textrm{spont}} =\left | \sum_{j=1}^NA_j \right |^2=T_d^{N-1}N^2.\label{eq:S3}
\end{equation}
A numerical evaluation (see Supplemental Fig.5 \cite{suppl}) of $\frac{\xi_{\textrm{stim}}}{\xi_{\textrm{spont}}}$ for  $T_d[\textrm{dB}]=0.88$ dB and $N\leq 5$ shows that the intensity of stimulated FWM scales slightly better than the spontaneous process (the maximum increase of the efficiency is $\sim12\%$ for $N=5$). 
On the other hand, the ratio between the scaling laws of the two processes exceeds $40\%$ (for $N=3$) in the experiment (Fig.\ref{Fig_4} of the main manuscript), indicating that the filtering losses 
\textcolor{black}{dominate }the drop 
\textcolor{black}{losses.}

\section*{Appendix D: Modeling the incoherent emission from the array}
\noindent In this section we derive an expression for the pair generation probability of the array as if the different rings were emitting incoherently and independently from each other. We also aim to cast this expression in a form 
\textcolor{black}{that} contains quantities 
accessible from the experiment. The starting point is Eq.(\ref{eq:S18}), where we set $I^{(N)}_{jk}=0$ for $j\neq k$ to remove the interference terms. In the experiment we do not have direct access to $B_j^{(N)}$, since the array geometry prevents \textcolor{black}{us from} 
individually pump\textcolor{black}{ing} source $j$ and \textcolor{black}{collecting pairs} 
from port $D_N$. 
What we \textcolor{black}{actually} can do is 
to pump source $j$ from port $D_j$ and collect photons from the Through port. This allows to \textcolor{black} {to} determine the pair generation probability immediately \emph{after} port $D_j$; we assume equal escape probabilities of photons in the Through and the Drop port, which finds \textcolor{black}{its} justification 
\textcolor{black}{in} the symmetric coupling of the ring with the two bus waveguides. From this, we need to compute the effective transmittance of the pair from source $j$ to port $D_N$, in order to correctly scale the pair generation probability and obtain $B_j^{(N)}$. Mimicking the experimental conditions, we considered a CW pump, which allows \textcolor{black}{us} to write Eq.(\ref{eq:S16}) in the form
\begin{multline}
    B_j^{\prime (N)} = \mathscr{K}^2\phi_P(\omega_{p_0})^2(-\gamma_{e,p})^{4(j-1)}|h_p(\omega_{p_0})|^{4j} \\ \int {\left |h_s^{N-j+1}(\omega_1)h_i^{N-j+1}(2\omega_{p_0}-\omega_1) \right |^2 d\omega_1}. \label{eq:S19}
\end{multline}
We now introduce the effective transmittance $T_{jN}$ of the pair from $j$ to $N$, 
\begin{equation}
    T_{jN} = \frac{\int \left |h_s^{N-j+1}(\omega_1)h_i^{N-j+1}(2\omega_{p_0}-\omega_1) \right |^2 d\omega_1}{\int \left |h_s(\omega_1)h_i(2\omega_{p_0}-\omega_1) \right |^2 d\omega_1}, \label{eq:S20}
\end{equation}
and insert it into Eq.(\ref{eq:S19}) to get
\begin{multline}
    B_j^{\prime (N)} = T_{jN} \left (\mathscr{K}^2\phi_P(\omega_{p_0})^2(-\gamma_{e,p})^{4(j-1)}|h_p(\omega_{p_0})|^{4j} \right. \\ \left. \int \left  |h_s(\omega_1)h_i(2\omega_{p_0}-\omega_1) \right |^2 d\omega_1 \right) \label{eq:S21}    
\end{multline}
The quantity in bracket\textcolor{black}{s} in Eq.(\ref{eq:S21}) is the pair generation probability of source $j$ when it is pumped from the input of the array, and when the emitted pairs are collected immediately \emph{after} port $D_j$. 
\textcolor{black}{D}enoting 
\textcolor{black}{by} $C_j$ the related coincidence rate, we have 
\begin{equation}
    \left(\frac{R_N}{R_q}\right)_{\textrm{incoh}} = \sum_{j=1}^N T_{jN}C_j. \label{eq:S22}
\end{equation}
This expression is used to estimate the normalized rate of incoherent emission from the array. In order to measure $C_j$, we injected light into port $D_j$, and scaled the power according to $P_j = P_1T_d^{j-1}$, where $T_d= (2\gamma_{e,p})\left |h_p(\omega_{p_0})\right |^2$ is the on-resonance drop transmittance. This emulates the $j-1$ drop events of the pump from the input to source $j$.

\nocite{bellegarde2018improvement,borghi2020phase}
%
\end{document}


\preprint{APS/123-QED}

\title{Supplemental material for Super spontaneous four-wave mixing  in an array of silicon microresonators}
\author{Massimo Borghi}
\email{Corresponding authors: massimo.borghi@unipv.it - matteo.galli@unipv.it}
\affiliation{Dipartimento di Fisica, Universit\`a di Pavia, Via Agostino Bassi 6, 27100 Pavia, Italy}

\author{Federico Andrea Sabattoli}\email{Current address: Advanced Fiber Resources Milan S.r.L, via Fellini 4, 20097 San Donato Milanese (MI), Italy}
\affiliation{Dipartimento di Fisica, Universit\`a di Pavia, Via Agostino Bassi 6, 27100 Pavia, Italy}

\author{Houssein El Dirani}
\email{LIGENTEC SA, 224 Bd John Kennedy, 91100 Corbeil-Essonnes, France}
\affiliation{Univ. Grenoble Alpes, CEA-Leti, 38054 Grenoble cedex, France}

\author{Laurene Youssef} \email{Current address: IRCER,   Centre Européen de la Céramique, 12 rue Atlantis, 87068 Limoges, France}
\affiliation{Univ. Grenoble Alpes, CNRS, LTM, 38000 Grenoble, France}



\author{Camille Petit-Etienne}
\affiliation{Univ. Grenoble Alpes, CNRS, LTM, 38000 Grenoble, France}

\author{Erwine Pargon}
\affiliation{Univ. Grenoble Alpes, CNRS, LTM, 38000 Grenoble, France}

\author{\textcolor{black} {J.}E. Sipe}
\affiliation{Department of Physics, University of Toronto, 60 St. George Street, Toronto, ON, M5S 1A7, Canada}

\author{Amideddin Mataji-Kojouri}
\affiliation{Dipartimento di Fisica, Universit\`a di Pavia, Via Agostino Bassi 6, 27100 Pavia, Italy}

\author{Marco Liscidini}
\affiliation{Dipartimento di Fisica, Universit\`a di Pavia, Via Agostino Bassi 6, 27100 Pavia, Italy}

\author{Corrado Sciancalepore}\email{Current address: SOITEC SA, Parc technologique des Fontaines, Chemin des Franques, 38190 Bernin, France}
\affiliation{Univ. Grenoble Alpes, CEA-Leti, 38054 Grenoble cedex, France}

\author{Matteo Galli} 
\affiliation{Dipartimento di Fisica, Universit\`a di Pavia, Via Agostino Bassi 6, 27100 Pavia, Italy}

\author{Daniele Bajoni}
\affiliation{Dipartimento di Ingegneria Industriale e dell'Informazione, Universit\`a di Pavia, Via Adolfo Ferrata 5, 27100 Pavia, Italy}




\date{\today}



\maketitle


\section{Experimental methods}
\label{Section: dev_fab}
\subsection{Sample fabrication}
\noindent The device was fabricated on a 200 mm Silicon\textcolor{black}{-on-Insulator} 
wafer with a 220 nm-thick device layer of crystalline silicon at CEA-Leti (Grenoble). Silicon structures are patterned by deep ultraviolet lithography with 120 nm resolution, followed by inductively coupled plasma etching (realized in collaboration with LTM (Laboratoire
des Technologies de la Microelectronique)) and $\textrm{O}_2$ plasma stripping. The sidewall roughness of the waveguides is reduced by Hydrogen annealing  \cite{bellegarde2018improvement}. After encapsulation resulting in a 900 nm-thick $\textrm{SiO}_2$ layer, 110 nm of titanium
nitride (TiN) 
\textcolor{black}{was} deposited and patterned to create the thermal phase
shifters. Aluminum copper (AlCu) was employed for the electrical pads.
A deep etch comprising two different steps ($\textrm{C}_4\textrm{F}_8/\textrm{O}2/\textrm{CO}/\textrm{Ar}$ plasma for the
whole depth of $\textrm{SiO}_2$, Bosch method to remove 150 $\mu$m of the 725 $\mu$m-thick Si
substrate) was implemented to separate the sub-dice and to realize high quality end facets for edge coupling.
\subsection{Device geometry}
\noindent The array in Fig.1(a) of the main text is made by $5$ ring resonators in the add-drop filter configuration. The cross section of the waveguides is $600\,\textrm{nm} \times 220\, \textrm{nm}$. Ring resonators are designed to be nominally identical, with a radius of $29.3\,\mu$m and a coupling gap with the bus waveguide of $120$ nm. The horizontal (center to center ) separation between the resonators in the array is $L=500\,\mu$m. 

\section{Linear characterization}
\begin{figure}[h!]
\centering\includegraphics[scale = 0.75]{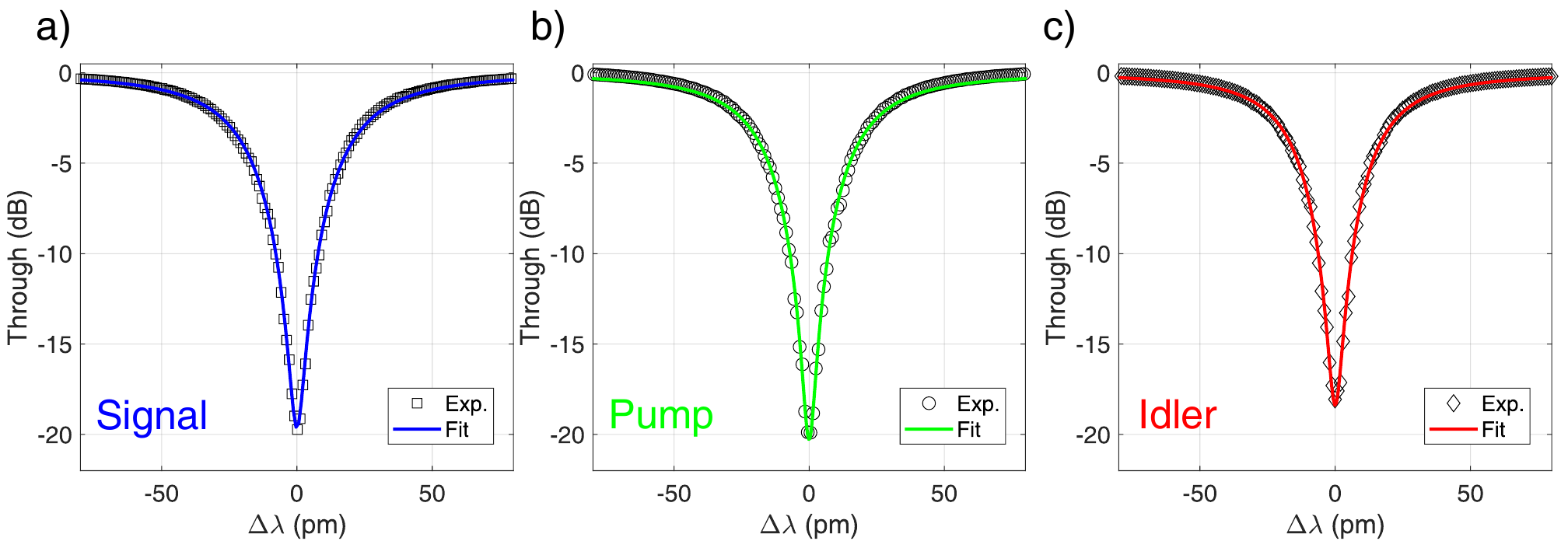}
\caption{(a) Spectral response of ring $1$ measured at the Through port of the resonator. The shift $\Delta\lambda$ is relative to the resonance wavelength of the signal $\lambda_s = 1571.2$ nm. (b) Same as in panel (a), but relative to the pump resonance at $\lambda_p = 1561.25$ nm. (c) Same as in panel (a), but relative to the idler resonance at $\lambda_i = 1551.4$ nm. In all the panels, scatters are experimental data, while solid lines are the best fit according to the model in Eq.(\ref{eq:S1}). \label{Fig_S2}}
\end {figure}

\begin{figure}[h!]
\centering\includegraphics[scale = 0.8]{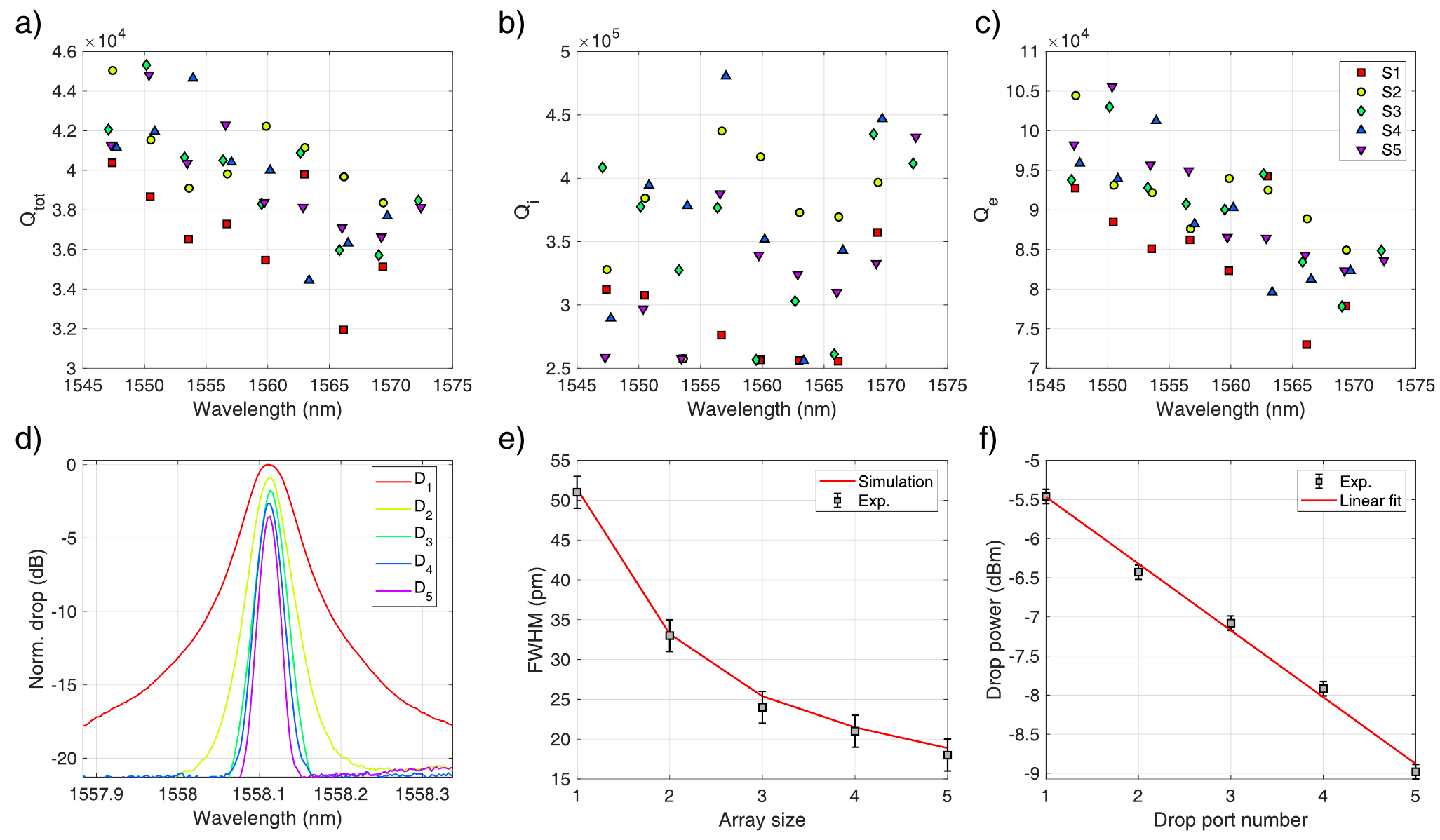}
\caption{(a) Total quality factor $Q_{\textrm{tot}}$ of the five sources $S_1-S_5$ as a function of wavelength. (b) Intrinsic quality factor $Q_{i}$ of the five sources as a function of wavelength. (c) Extrinsic quality factor $Q_{e}$ of the five sources as a function of wavelength. (d) Spectral response measured at the different drop ports $D_j$ of the array. (e) $3$-dB bandwidth of the drop transmittance shown in panel (d) as a function of the size of the array of resonators. Black scatters are experimental data, while the red line is the best fit predicted by simulation. The latter assumes identical rings with total and intrisic quality factors $Q_{\textrm{tot}}=3.9\times10^4$ and $Q_e=8.8\times10^4$. Those values have been determined from the experiment. (f) Power measured at the drop ports $D_j$ (see Fig.1(b) of the main text) when the pump light at $1561.25$ nm is resonantly injected at the input of the resonator sequence. The values at ports $D_2$ and $D_4$ have been corrected by taking into account the different transmittivity of the  collection fibers placed at the two opposite facets of the chip. The red line is a linear fit of the data. \label{Fig_S3}}
\end {figure}

\noindent The spectral response of each ring in the array is probed using a continuous wave (CW) tunable laser and a monitoring photodiode. The normalized transmittance $T_{H,m}(\omega)$ at the Through port around each resonance order $\omega_m$ is fitted according to the Temporal Coupled Mode Theory (TCMT) equation \cite{borghi2020phase},
\begin{equation}
    T_{H,m}(\omega) = \left | 1-\frac{2\gamma_e}{i(\omega-\omega_m)+\gamma_{\textrm{tot}}}\right |^2, \label{eq:S1}
\end{equation}
where $\gamma_e$ is the coupling rate between the ring and the waveguide and $\gamma_{\textrm{tot}}=2\gamma_{e}+\gamma_i$ is the total decay rate of the energy amplitude from the resonator. This term also accounts for the intrinsic loss rate $\gamma_i$ due to scattering of light off the chip. An example of the transmittance around the signal, the pump and the idler resonance order is shown in Fig.\ref{Fig_S2}. The fit based on Eq.(\ref{eq:S1}) is in very good agreement with the data. From Eq.(\ref{eq:S1}), we extracted the total $Q_{\textrm{tot}}=\frac{\omega_m}{2\gamma_{\textrm{tot}}}$, the extrinsic $Q_i=\frac{\omega_m}{2\gamma_i}$ and the extrinsic $Q_e=\frac{\omega_m}{2\gamma_e}$ quality factors for all the resonance orders in the wavelength interval $1545-1575$ nm, and for all the five resonator sources $S_1-S_5$. These are shown in Fig.\ref{Fig_S3}(a-c).

\begin{table}[h!]
\begin{centering}
\begin{tabular}{c|c|c|c}
Parameter & $\lambda_{i}=1551.4$ nm & $\lambda_{s}=1571.2$ nm & $\lambda_{p}=1561.25$ nm \\
\hline 
$Q_{\textrm{tot}}$ & $(4.2\pm0.3)\times10^{4}$ & $(3.7\pm0.1)\times10^{4}$ & $(3.9\pm0.2)\times10^{4}$ \\ 
\hline 
$Q_{i}$ & $(3.5\pm0.5)\times10^{5}$ & $(3.9\pm0.5)\times10^{5}$ & $(3.2\pm0.7)\times10^{5}$\\
\hline 
$Q_{e}$ & $(9.7\pm0.7)\times10^{4}$ & $(8.1\pm0.3)\times10^{4}$ & $(8.8\pm0.4)\times10^{4}$\\ \hline
FSR (nm) & $(3.102\pm0.002)$ & $(3.188\pm0.002)$ & $(3.154\pm0.004)$  \\
\end{tabular}
\end{centering}
\caption{Average values of the total ($Q_{\textrm{tot}}$), intrinsic ($Q_i$) and extrinsic ($Q_e$) quality factors of the rings which compose the array. \label{Tab_1}}
\end{table}
\noindent Their values, averaged over all the rings in the array, is reported in Table \ref{Tab_1}. The table also reports the free spectral range (FSR), whose variation ($\sim 3$ pm) among the rings is an order of magnitude lower than their resonance linewidth ($\sim 50$ pm). The great uniformity in the FSR and in the total quality factors \textcolor{black}{allows} to perfectly overlap the spectra of the different resonators. This is accomplished by regulating the current in the metallic heaters. Once overlapped, we measured the spectra at the output ports $D_j$ (see Fig.1(b) in the main text for a description of the ports location), which are shown in Fig.\ref{Fig_S3}(d). When the device is probed at port $D_j$, the transmission spectra of all the rings from $1$ to $j$ are overlapped, while ring $j+1$ is detuned by more than three times its resonance linewidth from the others to make it transparent to the incoming light. From the spectra in Fig.\ref{Fig_S3}(d), we extracted the Full Width at Half Maximum as a function of the size of the array, and reported it in Fig.\ref{Fig_S3}(e). As expected, the bandwidth progressively narrows as we add more resonators to the sequence, due to the combined filtering effect of each element in the array. This trend is well reproduced by a numerical simulation where rings are assumed to be all identical and with an overlapped spectrum, as shown by the solid red line in Fig.\ref{Fig_S3}(e). \\
\noindent As a last step, we determined the on-resonance transmittance $T_d$ associated to each drop event in the sequence. To this end, we resonantly coupled the CW laser to the input of the first ring, with the wavelength set to $\lambda=\lambda_i=1551.4$ nm, and we measured the scaling of the pump power at the output of the different drop ports $D_j$. This is shown Fig.\ref{Fig_S3}(f), after accounting for the different transmittivity of the lensed fibers to the left and to the right hand side of the chip (see the experimental setup in Fig.1(c) of the main manuscript). From a linear fit of the data, we extracted a transmittance of $(T_d[\textrm{dB}] = 0.88\pm0.07)$ dB for each drop event. This value is compatible with that predicted by TCMT\textcolor{black}{,} $T_{d,\textrm{TCMT}}[\textrm{dB}]=20 \textrm{log} \left ( \frac{Q_{\textrm{tot}}}{Q_e}\right ) = (-1.1\pm0.2) $ dB, where the values of $Q_{\textrm{tot}}$ and $Q_i$ have been taken from Table \ref{Tab_1} (idler wavelength). From TCMT, we also have a predicted extinction ratio of $\textrm{ER}_{\textrm{TCMT}}[\textrm{dB}] =20 \textrm{log} \left (1-\frac{Q_{\textrm{tot}}}{Q_e}\right ) = (-18\pm 1) $ dB, which is compatible with the measured value of $\textrm{ER}[\textrm{dB}]=-18.2\pm0.2$ dB (see Fig.\ref{Fig_S2}(c)).
\newpage
\section{Supplemental figures}
\begin{figure}[h!]
\centering\includegraphics[scale = 0.7]{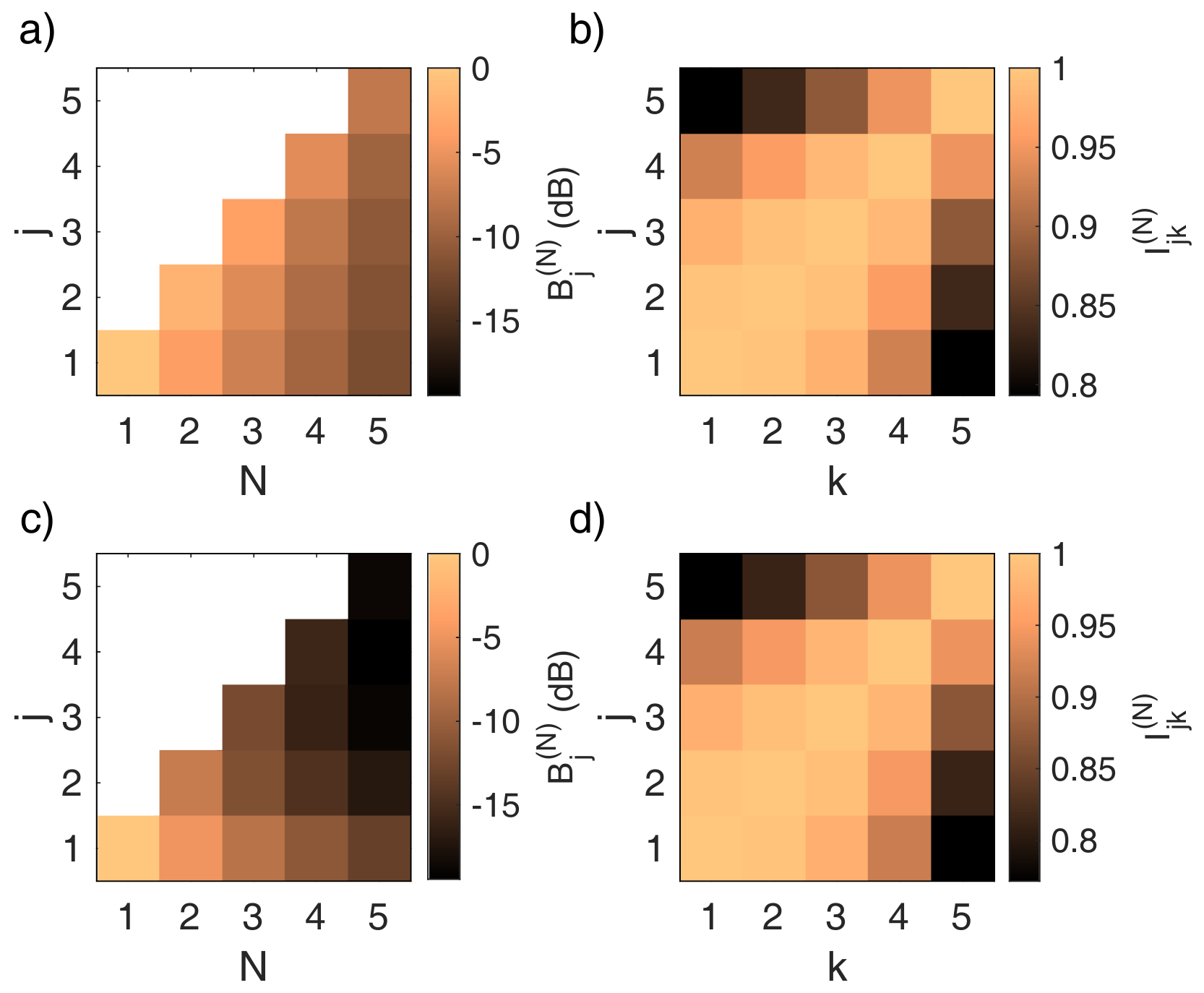}
\caption{(a) Relative brightness $B^{(N)}_j$ (compared to a single source) of source $j$ as seen from the output of a sequence of $N$ resonators. The calculation assumes a CW excitation. (b) Indistinguishability $I^{(N)}_{jk}$ between the JSA of source $j$ and $k$ as seen from the output of a sequence of $N$ resonators. The calculation assumes a CW excitation. (c) Same as in (a), but relative to a pulsed excitation which uses the pump spectral amplitude shown in Fig.\ref{Fig_pump}. (b) Same as in (b), but relative to a pulsed excitation. \label{Fig_eff_trasm}}
\end {figure}

\begin{figure}[t!]
\centering\includegraphics[scale = 0.7]{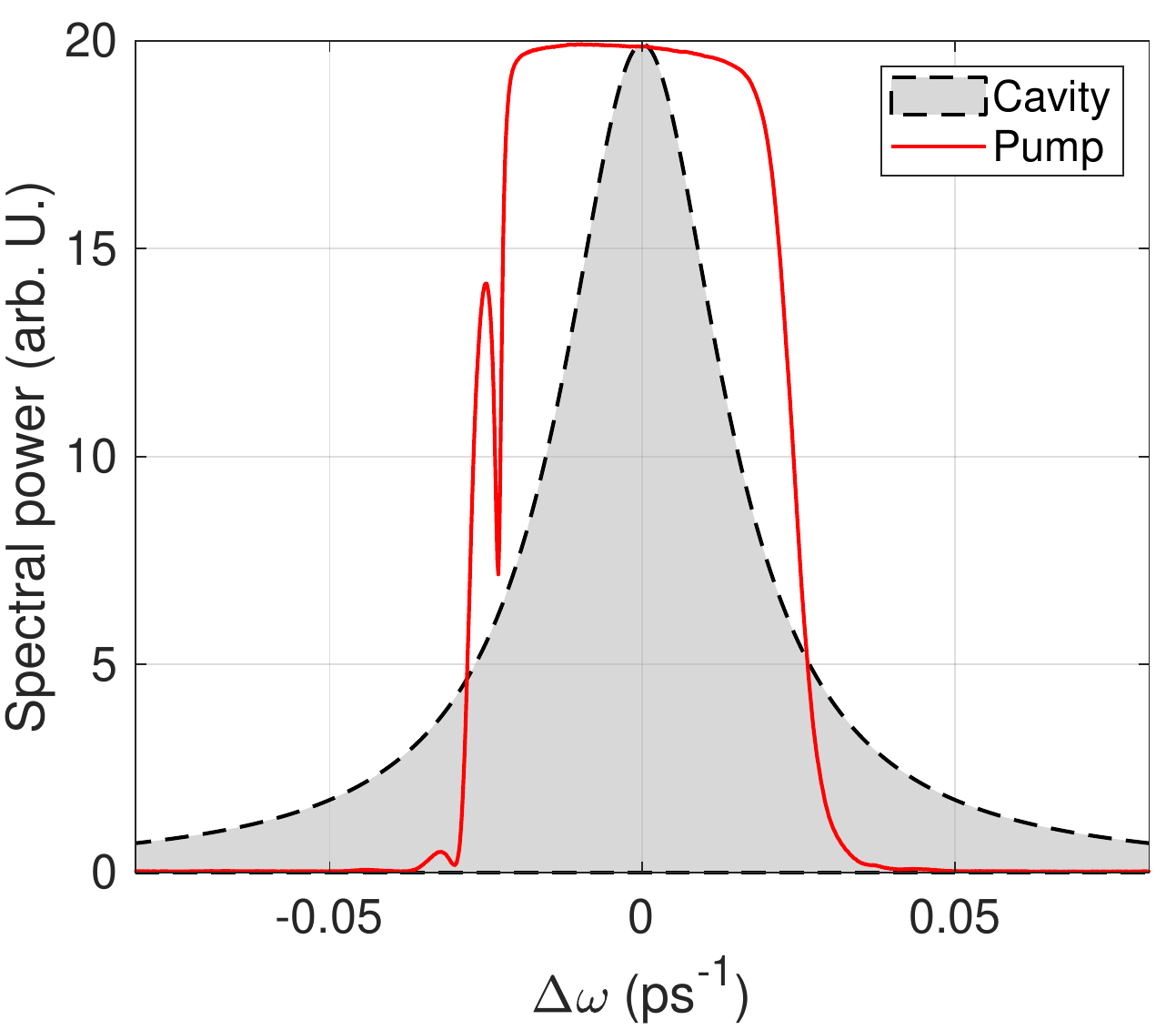}
\caption{Measured power spectral amplitude of the pulsed laser (red). A simulation of the spectral response of  the resonators is superimposed in gray for comparison. \label{Fig_1}\label{Fig_pump}}
\end {figure}

\begin{figure}[t!]
\centering\includegraphics[scale = 0.62]{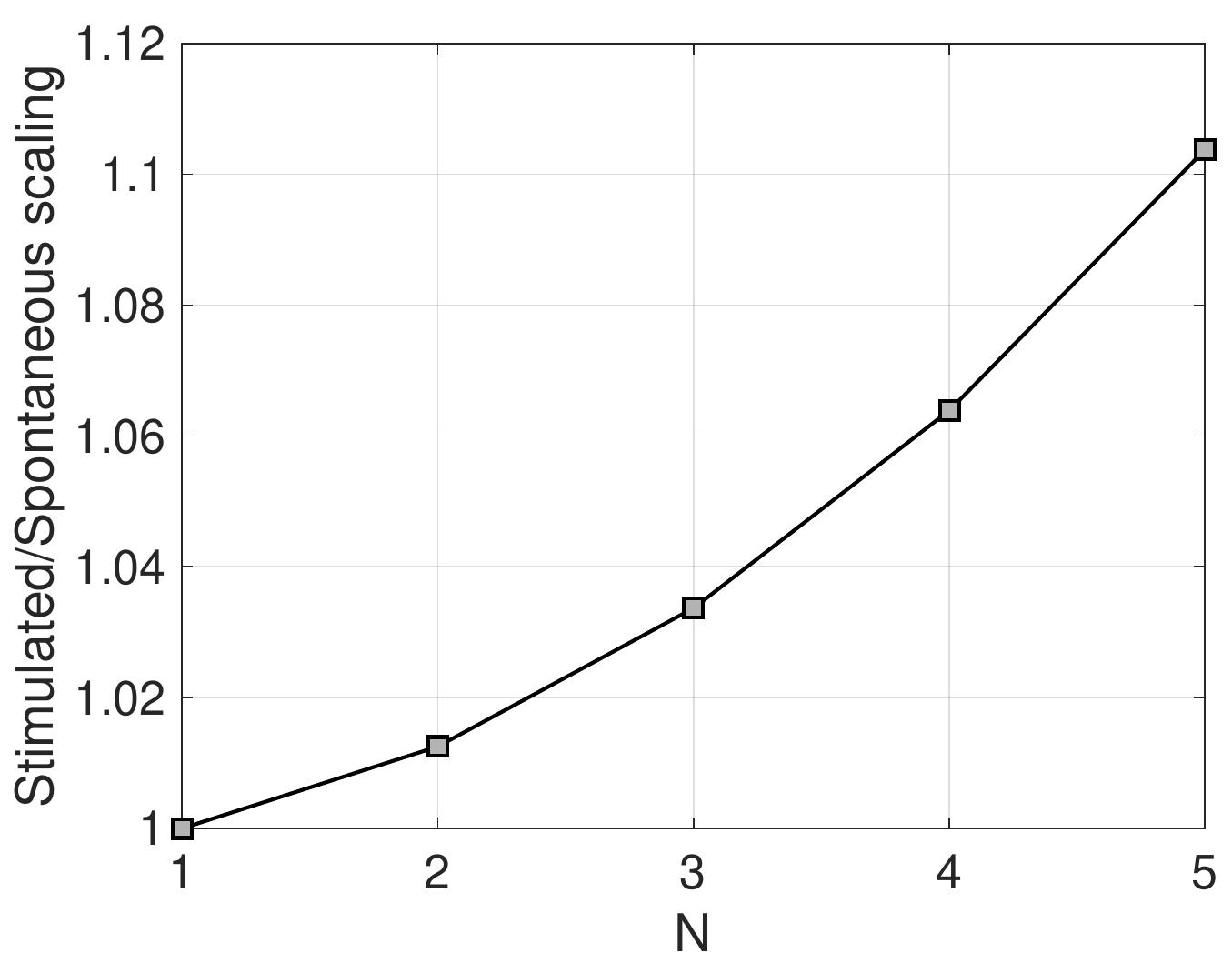}
\caption{The ratio $\frac{\xi_{\textrm{stim}}}{\xi_{\textrm{spont}}}$ between the scaling laws of stimulated and spontaneous FWM in absence of spectral filtering. \label{Fig_S4}}
\end {figure}
%
